\documentclass{wlscirep}

\usepackage{PRIMEarxiv}
\usepackage[utf8]{inputenc} %
\usepackage[T1]{fontenc}    %
\usepackage{hyperref}       %
\usepackage{url}            %
\usepackage{booktabs}       %
\usepackage{amsfonts}       %
\usepackage{nicefrac}       %
\usepackage{microtype}      %
\usepackage{lipsum}
\usepackage{fancyhdr}       %
\usepackage{graphicx}       %
\usepackage{subcaption}
\usepackage{caption}       %
\graphicspath{{figures/}}     %
\usepackage{colortbl}
\usepackage{placeins}
\usepackage{soul,color}
\usepackage{amssymb}
\usepackage{cite}
\usepackage{tabularx}
\usepackage{xspace,amstext}
\usepackage{amsmath}
\usepackage{adjustbox}
\usepackage{comment}
\usepackage{tikz}
\usepackage{listofitems} 
\tikzset{>=latex} %
\usepackage{xcolor}
\colorlet{myred}{red!80!black}
\colorlet{myblue}{blue!80!black}
\colorlet{mygreen}{green!60!black}
\colorlet{myorange}{orange!70!red!60!black}
\colorlet{mydarkred}{red!30!black}
\colorlet{mydarkblue}{blue!40!black}
\colorlet{mydarkgreen}{green!30!black}
\tikzstyle{node}=[thick,circle,draw=myblue,minimum size=22,inner sep=0.5,outer sep=0.6]
\tikzstyle{node in}=[node,green!20!black,draw=mygreen!30!black,fill=mygreen!25]
\tikzstyle{node hidden}=[node,blue!20!black,draw=myblue!30!black,fill=myblue!20]
\tikzstyle{node out}=[node,red!20!black,draw=myred!30!black,fill=myred!20]
\tikzstyle{connect}=[thick,mydarkblue] %
\tikzset{ %
  node 1/.style={node in},
  node 2/.style={node hidden},
  node 3/.style={node out},
}
\def\nstyle{int(\lay<\Nnodlen?min(2,\lay):3)} %

\usepackage{tkz-graph}

\usetikzlibrary{matrix}

\usepackage{lineno}

\usepackage{lipsum}
\title{ Deep Learning-Based Classification of Gamma Photon Interaction in Room-Temperature Semiconductor Radiation Detectors }

\author[1,$ \dag  $]{Sandeep K. Chaudhuri}
\author[2,$ \dag  $]{Qinyang Li}
\author[1,$ \ddag  $,+]{Krishna C. Mandal}
\author[2,$ \ddag  $,*]{Jianjun Hu}

\affil[1]{Department of Electrical Engineering\\
  University of South Carolina\\
    Columbia, SC 29201 }
\affil[2]{Department of Computer Science and Engineering\\
      University of South Carolina              \\
  Columbia, SC 29201  }
  
\affil[$ \dag  $]{these authors contributed equally to this work}
\affil[$ \ddag  $]{Corresponding Author}
\affil[+]{\texttt{mandalk@cec.sc.edu}}
\affil[*]{\texttt{jianjunh@cse.sc.edu}}

\begin{document}
\maketitle

\begin{abstract}

Photon counting radiation detectors have become an integral part of medical imaging modalities such as Positron Emission Tomography or Computed Tomography. One of the most promising detectors is the wide bandgap room temperature semiconductor detectors, which depends on the interaction gamma/x-ray photons with the detector material involves Compton scattering which leads to multiple interaction photon events (MIPEs) of a single photon. For semiconductor detectors like CdZnTeSe (CZTS), which have a high overlap of detected energies between Compton and photoelectric events, it is nearly impossible to distinguish between Compton scattered events from photoelectric events using conventional readout electronics or signal processing algorithms. Herein, we report a deep learning classifier CoPhNet that distinguishes between Compton scattering and photoelectric interactions of gamma/x-ray photons with CdZnTeSe (CZTS) semiconductor detectors. Our CoPhNet model was trained using simulated data to resemble actual CZTS detector pulses and validated using both simulated and experimental data. These results demonstrated that our CoPhNet model can achieve high classification accuracy over the simulated test set. It also holds its performance robustness under operating parameter shifts such as Signal-Noise-Ratio (SNR) and incident energy. Our work thus laid solid foundation for developing next-generation high energy gamma-rays detectors for better biomedical imaging.

. %

\end{abstract}

\keywords{CdZnTeSe \and deep learning \and deep learning one-class classifier \and gamma-photon detection \and graph neural network \and medical imaging \and multiple interaction photon events\and photon counting detection \and radiation detection\and semiconductor detectors}

\section{Introduction}

While X-ray detection has been historically used for radiology and Computed Tomography (CT) of organs in living species, gamma-ray imaging modalities such as positron emission tomography (PET) or Single Photon Emission Computed Tomography (SPECT) are crucial to medical diagnosis of complex diseases like cancer, detailed imaging of internal structures, organs, and bones, and functional imaging of vital organs such as heart, brain, thyroid, and lungs \cite{Jaffray2023, Stein2023, Fontenele2023, Skrzynski2022, Sheng2023}. The relatively higher energy of gamma-rays compared to x-rays makes them more penetrable in tissues and bones, hence are particularly useful for deep-seated structures or small abnormalities within organs under investigation. While both x-rays and gamma-rays are forms of ionizing energy radiation and capable of causing cellular damage upon overexposure, optimization of dosage or exposure to gamma rays is of particular importance due to their higher energies \cite{Hosono2021}.

The dosage optimization is largely limited by the response of the radiation detectors used to detect the photons transmitted through the organs. Due to the slower response, poor energy and spatial resolution, conventional detectors such as film-screen radiography, scintillators, or solid-state photoconductors require longer exposure times. Semiconductor detectors are particularly advantageous for photon counting detection (PCD) as they are direct readout, require low power for operation, and provide high spatial resolution, energy resolution, efficiency, sensitivity, and faster response time which conventional detectors cannot provide \cite{Enlow2023, Rajendran2021, Hsieh2020, Takeuchi2016}. PCD is also used to monitor special nuclear materials and spent nuclear fuel as a measure for nuclear non-proliferation, and radiation leakage in and around nuclear reactors as a measure of radiation safety.  

Semiconductor detectors can be easily integrated into hybrid imaging systems, combining different imaging modalities (such as PET/CT or SPECT/CT) allowing for obtaining comprehensive anatomical and functional information from a single imaging session \cite{Takeuchi2016}. Semiconductor detectors are less susceptible to many artifacts present in conventional imaging techniques such as scintillation light scattering. Due to their higher material density, semiconductor detectors have much higher stopping power for penetrating gamma photons. Among semiconductor imaging detectors, a wide bandgap compound semiconductor CdZnTe (CZT), stands out because of its high atomic number (Z) constituents and room-temperature operability \cite{Kabir2017, Niraula2013, Veale2023, Chaudhuri2022}. Being a high-Z semiconductor, CZT stops gamma-rays much more effectively than conventional silicon or silicon carbide detectors \cite{Mandal2023, Mandal2013} and because of its wide bandgap ($>1.6$ eV) CZT detectors can be operated at room-temperature and above unlike silicon or germanium x-/gamma-ray detectors. As a result, CZT detectors have become mainstream imaging detectors and are commercially fabricated for SPECT, CT, and PET imaging systems for diagnostic applications in oncology, cardiology, bone densitometry, and dentistry \cite{Kromek2023}.

CZT detectors are albeit expensive due to poor crystal growth yield (on the order of 30\%) which makes the imaging system as well as the end user expenses higher. A recently discovered quaternary semiconductor CdZnTeSe (CZTS) offers wide bandgap (1.6 eV), excellent charge transport properties, and crystal growth yield above 90\% \cite{Roy2019}. With all the desired properties of a room temperature gamma-detector and such high growth yield, CZTS is poised to be the next-generation semiconductor detector for economical yet high-resolution high-contrast digital imaging systems \cite{Roy2021, Chaudhuri2023, Kleppinger2023}.

\begin{figure}[ht]
  \centering
  \includegraphics[width=0.6\linewidth]{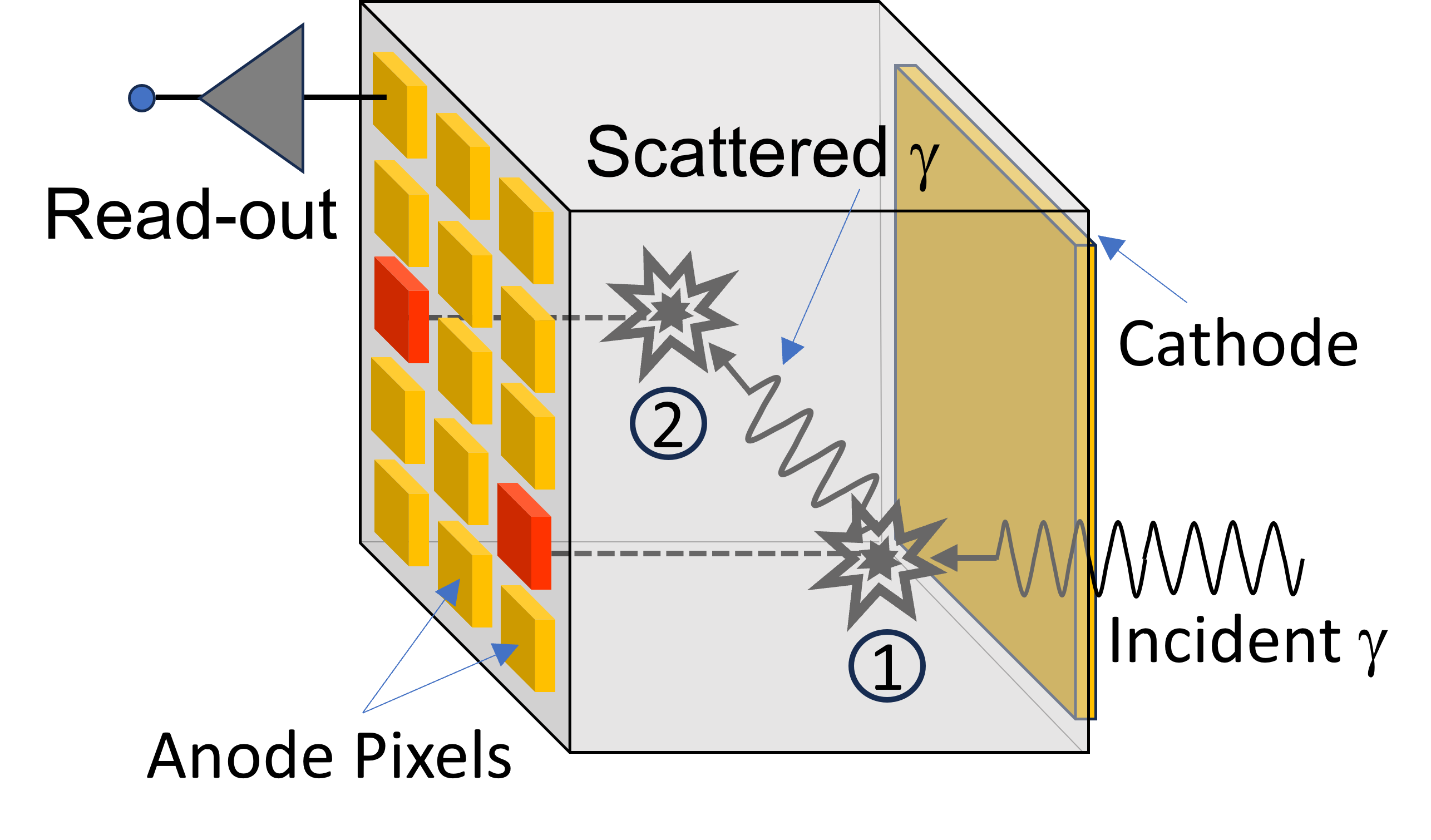}
    \caption{CZTS gamma detector with CS/PE pulses. A Compton scattered event (1) followed by a photoelectric event (2) in a multiple interacion photo event (MIPE) resulting in the triggering of two anode pixels shown in red for a pixelated detector. (Note: the readout electronic block is shown for a random pixel only).}
    \label{fig:pixelated}
\end{figure}

In spite of the above advantages, CZTS (or CZT) detectors have inherent limitations due to the way x-/gamma- photons interact with the detector. Gamma-rays with energies below 1 MeV typically interact with matter through either photoelectric (PE) interactions or Compton scattering (CS) (Fig. \ref{fig:pixelated}). PE interactions are preferred in PCD as the gamma photons deposit their entire energy in a single interaction. PCD works on the principle of measuring the energy of the individual photons interacting with the detectors. PCD often employs pixelated detectors which contain a large number of closely spaced pixels to obtain high spatial resolution. In the case of CS, a single gamma photon interacts with a detector multiple times causing multiple interaction photon events (MIPEs) \cite{Levin167, Gu2010, Farahmandzadeh1514}. CS interactions leads to partial energy deposition causing massive loss in spectral information. CZTS (or CZT) detectors have high mass attenuation coefficient for Compton scattered gamma events leading to a significant fraction of MIPEs. Due to the MIPEs the detectors record a series of energy depositions, as shown in Fig. \ref{fig:pixelated}, that does not correspond to the original energy of the photon resulting in false triggering of multiple pixels, hence resulting in poor image resolution and contrast.

\begin{figure}[!htb] 
  \vspace{-0.5em}
  
  \begin{subfigure}[b]{0.49\textwidth}
  \centering
    \includegraphics [width=1\textwidth]{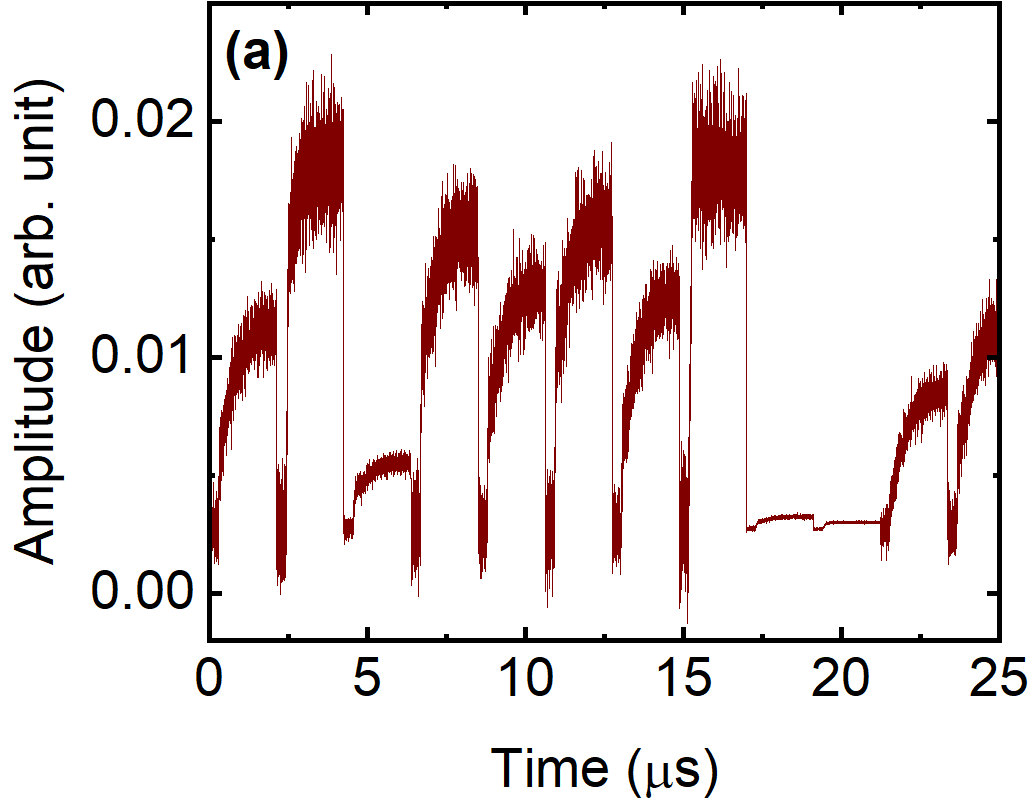}
  \end{subfigure}
\hspace{1.5em}
  \begin{subfigure}[b]{0.49\textwidth}
    \centering
    \includegraphics [width=1.\textwidth]{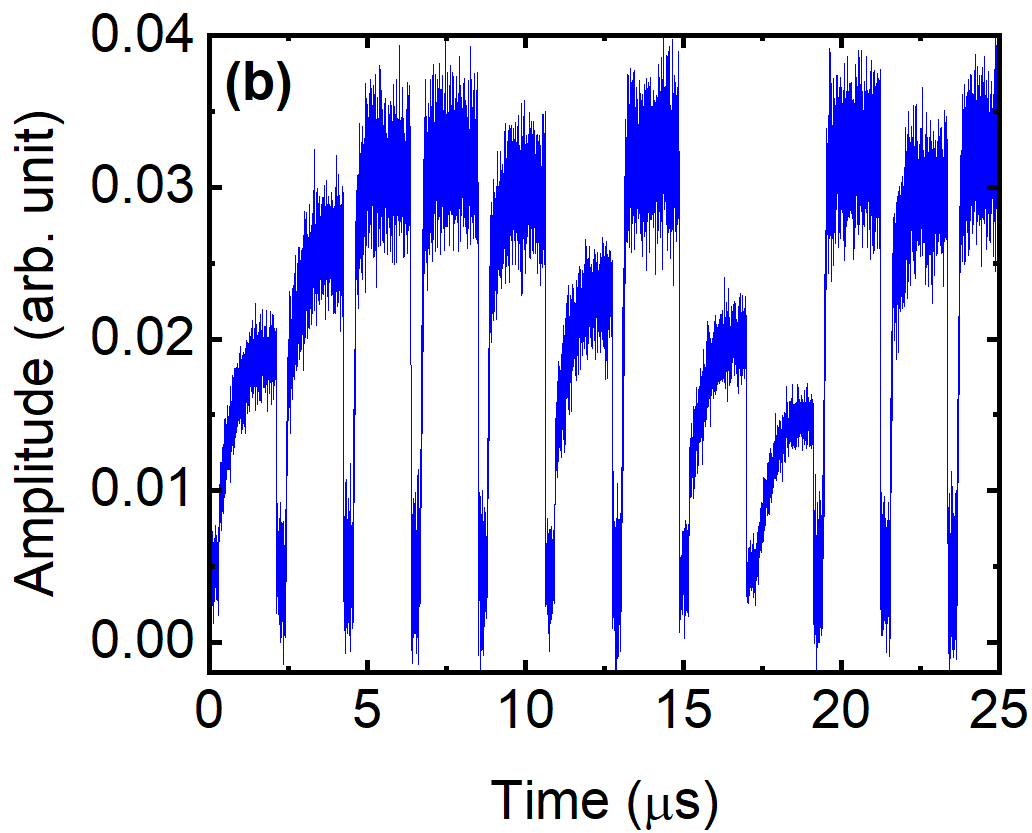}
  \end{subfigure}

  \caption{Sample CS (a) and PE (b) pulse signal trains. It is non-trivial to distinguish these two types of signals based on their shape.
  }
  \label{fig:pulsetrains}
\end{figure}

An energy discrimination circuit may be used in an ideal detection system to reject CS events as they register lower energies than any PE event. However, in CZTS (or CZT) detectors there is a significant overlap of deposited energies among the two types of interactions. In such situations discrimination of events based on energies may lead to significant decline of count rate leading to increased dosage in patients. In practical CZTS detectors, electronic circuits or any conventional pulse shape analysis algorithm cannot distinguish a CS event from a PE event if they result in detector output signals with same pulse heights. Fig. \ref{fig:pulsetrains} (a) and (b) shows pulse trains generated for a series of CS and PE events to demonstrate that distinguishing between these two types of interactions is non-trivial based on pulse-shape alone.

Furthermore, as previously discussed, conventional algorithms struggle to effectively differentiate CS events from PE events, prompting the adoption of deep learning neural network classifiers, which have the capacity to autonomously learn hierarchical data representations, uncovering intricate patterns and features across various levels of abstraction—a crucial advantage for our classification task.
However, neural networks operate as black boxes, yielding only final predicted labels for input data, devoid of interpretable knowledge or patterns for human experts to analyze. 

In this article, we report the development of a deep learning (DL) model, Compton-Photoelectric-neural-network (CoPhNet), which classifies CS events and PE events generated due to the interactions of monoenergetic (662 keV) gamma-photons with CZTS detectors with a high degree of confidence. The CoPhNet model is trained using charge pulses simulated for CS and PE events separately for a CZTS detector-preamplifier assembly with response resembling an actual detection system. Each training dataset comprises tens of thousands of pulses containing 5 million data points. Then the model has been validated using a simulated dataset with randomly mixed CS and PE events. Another challenge we face is the difficulty to evaluate the performance over experimental data, since it is infeasible to access the ground truths of the pulse labels in experimental data.  Inspired by the approach detailed in \cite{doi:10.1021/acs.jcim.3c00224}, where the authors harnessed t-Distributed Stochastic Neighbor Embedding (t-SNE), a potent dimensionality reduction technique, to visualize and explore complex high-dimensional data, particularly internal feature vectors of neural networks, we adopted a similar principle. By applying this method, we gain the ability to illustrate the distribution of the unlabeled dataset, enhancing the interpretability and robustness of our results. The described approach is poised to revolutionize the field of radiation imaging by addressing the above crucial limitations in present day imaging systems as well as in applications such as radiation spectroscopy, materials science, radiation dosimetry, nuclear physics, and environmental monitoring that needs to distinguish Compton events from photoelectric interactions.

\section{Method}
\label{sec:Method}

Our work flow for the training and the validation of the CoPhNet
is as follows:
\begin{itemize}
        \item    Generate the datasets and prepare the training and test sets
        \item    Train the neural network model using the training set
        \item    Feed the test set to the trained network to get predicted labels and evaluate its performance over simulated datasets
        \item    Visualize the feature vectors to understand its performance and evaluate the performance over the experimental dataset
\end{itemize}
\subsection{Dataset preparation and generation}

The detector pulses for training have been generated through a Monte-Carlo simulation algorithm coded in MATLAB. The algorithm has been specifically designed to take into account the effects of charge trapping and the front-end readout electronics, such as preamplifier to resemble an actual CZTS detection system.

\begin{figure}[ht]
  \centering
 \includegraphics[width=10cm]{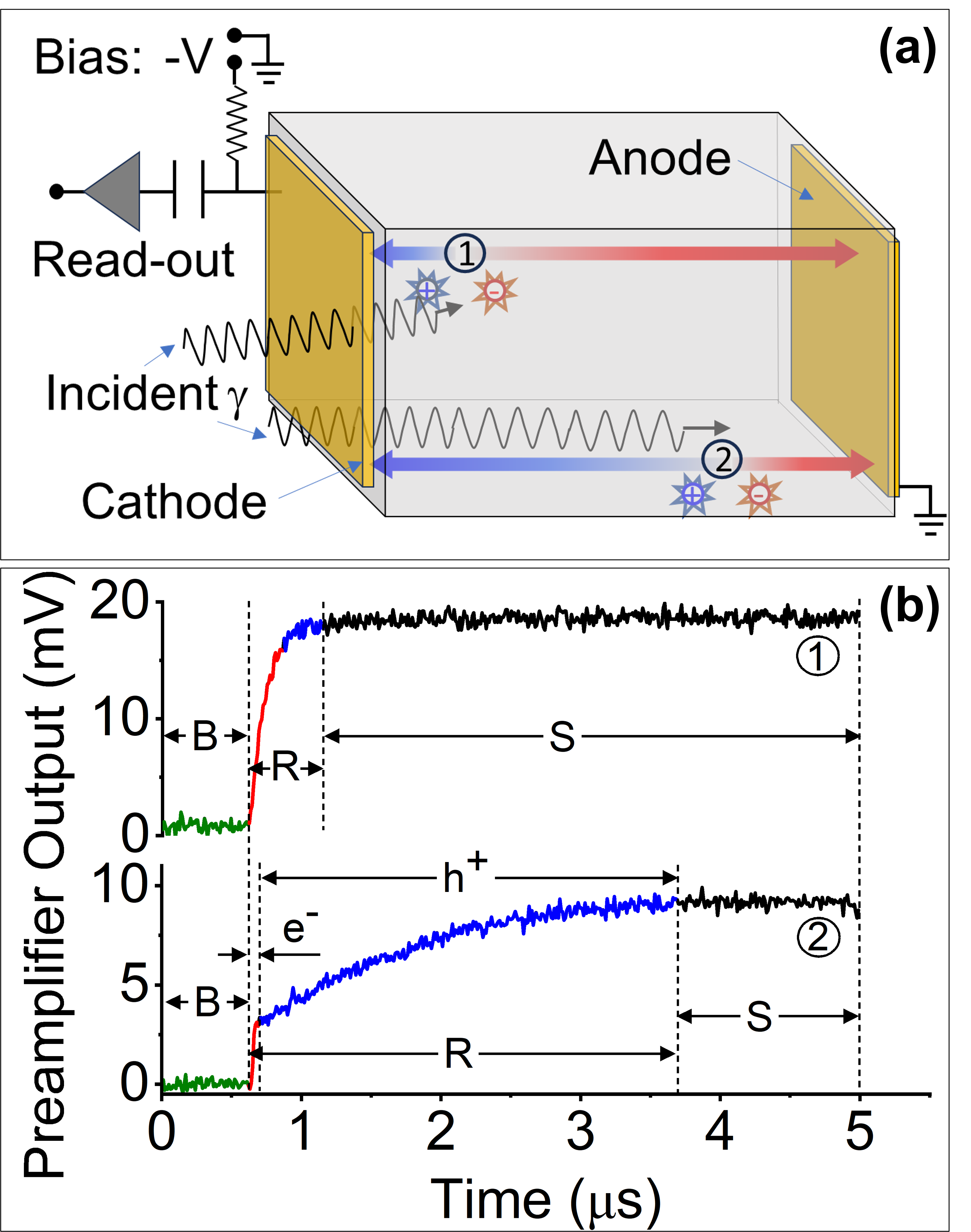}
    \caption{a) Two photon interaction events, one close to the cathode (1) and other close to the anode (2) in a planar detector with the collecting electrode biased at a negative potential versus anode. b) The pulses at the preamplifier  output corresponding to the situations (1) and (2), respectively. For the case of (1), the electron (e$^-$) transport predominantly forms the detector signal while the holes (h$^+$) with relatively poor charge transport property forms the signal in the case of (2) leading to a much longer rise time (B: Baseline; R: Rise; S: Saturation).}
    \label{fig:signals}
\end{figure}

As mentioned above, the energy deposited by the incident photon in the detector in PCD is measured for each interaction. When a photon interacts with the detector it generates a number of electron-hole charge pairs proportional to the photon energy. The generated charge pairs form the detector signal while transiting or drifting along the detector thickness under the influence of an applied bias. In a typical detection system the detector is connected to a preamplifier that generates a voltage pulse for each interaction, with a pulse-height proportional to the number of electron-hole pairs (ehp) generated by the interaction. Hence, by measuring the individual preamplifier pulse-height the incident photon energies can be determined for each interaction. Fig. \ref{fig:signals} (a) summarizes the schematic of a typical PCD system illustrated for a planar detector.

Fig. \ref{fig:signals} (b) shows typical charge pulses, plotted as the evolution of the voltage as a function of time, with three distinguishable parts: i) the baseline defines the reference point of the measured voltage, ii) the rising corresponds to the electron-hole pair drift under the influence of the applied bias, iii) a flat top corresponding to the saturation charge collected by the detector electrodes. The pulse height is the voltage difference between the baseline and the flat top. However, in practice, the pulse height is measured after filtering the electronic noise which in our case is a semi-Gaussian (RC-CR$^4$) shaping algorithm \cite{Nakhostin2012, Chaudhuri2012}. 

The rising part can be seen to have two distinctive slopes (rise time) which arises due to the difference in electron and hole transport properties in CZTS and other room temperature semiconductor detectors. The faster rise corresponds to the electron transit and the slower one corresponds to the hole transit. Hole transport is severely impeded in CZTS compared to that of electrons due to the presence of excessive hole trapping defect centers \cite{chaudhuri2021quaternary, ChaudhuriJAP2020}. For interactions close to the anode the charge transport is mostly due to the electrons while for those close to the cathode the hole transport dominates. Gamma rays interact at random thickness within the detector and depending on the position of the interaction the average rise times of the pulses vary. 

\paragraph{Experimental dataset generation}

\begin{figure}[ht]
  \centering
  \includegraphics[width=0.8\linewidth]{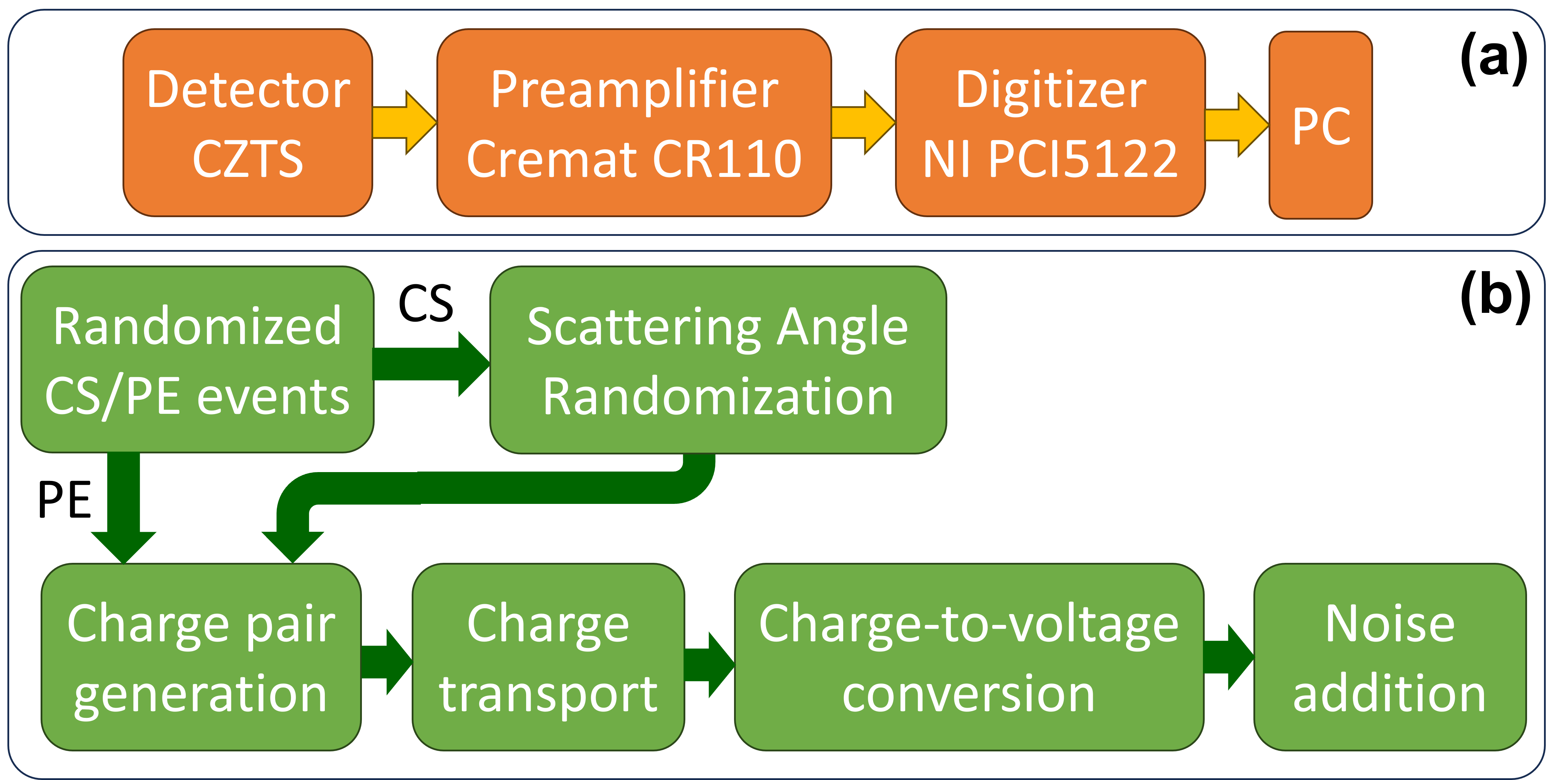}
    \caption{(a) Schematic of the acquisition set up for obtaining the experimental data using a CZTS detector. (b) Block diagram explaining the flow of the pulse simulation code.}
    \label{fig:setup}
\end{figure}

The experimental dataset was obtained using a CZTS detector with the characteristics mentioned in Table \ref{tab:para}. Fig. \ref{fig:setup} (a) shows the schematic of the experimental set up. A $^{132}$Cs radioisotope emitting monoenergetic (662 keV) gamma rays was used to illuminate the detector biased at 100 V. The detector was connected to a Cremat CR110 charge sensitive preamplifier. The output of the preamplifier was digitized using a NI PCI5122 fast digitizer card interfaced using a LabVIEW based data acquisition code. Details of the digital data acquisition system may be found elsewhere \cite{ChaudhuriJAP2020}

\begin{table}[h]
\begin{center}
\caption{ Detector parameters used for data generation. }
\label{Material specs}
\begin{tabular}{|l|l|l|l|}
\hline
Electron mobility-lifetime ($\mu\tau$) product & 2.9 $\times$ 10$^{-3}$ cm$^2$/V \\ \hline
Hole mobility-lifetime ($\mu\tau$) product & 1.4 $\times$ 10$^{-4}$ cm$^2$/V \\ \hline
Electron drift mobility & 760 cm$^2$/V.s \\ \hline
Hole drift mobility & 40 cm$^2$/V.s \\ \hline
Detector thickness & 0.16 cm \\ \hline
Material density & 5.8 cm$^-3$ \\ \hline
Applied bias & 100 V \\ \hline
\end{tabular}
\label{tab:para}
\end{center}
\end{table}

\paragraph{Simulation dataset generation}

Fig. \ref{fig:setup} (b) shows the schematic of the pulse simulator. The code first generates the random location of interactions which are calculated based on the attenuation coefficients obtained from the NIST's XCOM photon cross-section database \cite{berger2010xcom}. A planar detector with a detector thickness of 1.6 mm and an applied bias of 100 V has been considered. Next, the number of ehps generated in each interaction were calculated assuming an ehp creation energy of 4.67 eV/ehp. The ehp creation energy has been assumed to be equal to that in CZT as both have similar bandgap energy and crystal structure. For the CS events, the deposited energy depends on the angle of scattering. For this purpose, randomly generated angles of scattering were used to calculate the deposited energy using (1) which in turn were used to calculate the number of ehps generated. Following the calculation of the number of ehps, the charges induced on the collecting electrode were calculated as a function of position using the Shockley-Ramo-Nash theorem for charge transport in a planar detector geometry. The effect of charge trapping has been taken into consideration while calculating the drift velocities and drift times using the mobility-lifetime ($\mu\tau$) products and drift mobilities of electrons and holes calculated for a real CZTS detector. The detector parameters have been provided in Table \ref{tab:para}. A conversion factor of 1.4 $\times$ 10$^{12}$ V/C resembling that of a CR110 preamplifier was used to express the induced charge in Volts. Next, a baseline with a duration similar to our detection system has been added to the charge pulses. Finally, a Gaussian white noise was added to the signals to obtain a signal-to-noise ratio similar to the actual detection system. A series of 10,000 pulses each containing 500 data points with a sampling time of 1 $\times$ 10$^{-8}$ s consisted of a dataset. For the training, separate datasets were generated for pure CS and PE events. 
For the validation purpose, datasets with randomly mixed CS and PE events were generated.

Radiological imaging uses different radioisotopes depending on the imaging requirement which emits x-/gamma photons of varied energies. Similarly, in nuclear environments such as storage sites or reactors, gamma photons with various energies are emitted. Therefore, a practical requirement for the detection system would be that the classifier algorithm, although trained for one specific gamma energy, should still be able to perform with high accuracy for incident gamma-photons of various energies. As a result our CoPhNet, while trained with 662-keV gamma-photon interaction dataset, has been validated for datasets simulated for gamma photons in the energy range 1 keV - 1 MeV. Similarly, a real imaging system, of different noise levels, comprises many detector elements or pixels. Hence, the CoPhNet when trained for a detector with a particular noise level should be evaluated for the situations where a detector has a noise level different from that used for the training purpose. The CoPhNet has been validated for simulated datasets comprising with different noise levels while trained with a dataset pulses with a particular noise level.  
All the validation datasets' simulation parameters are mentioned in Table \ref{tab:all_sim}.

\begin{table}[!htb]
\begin{center}
\caption{ Description of simulated and experimental datasets for training and testing. }
\label{Data specs}
\begin{adjustbox}{max width=1.1\textwidth,center}
\begin{tabular}{|m{2.5cm}|m{2cm}|m{1.5cm}|l|m{2cm}|m{2cm}|l|}
\hline
Data type & Interaction type & Incident energy (keV) & No. of pulses & Sampling Interval (s) & Sampling points/pulse & SNR (dB) \\ \hline
Simulated data for training & CS & 662 & 8,000 & 1 $\times$ 10$^{-8}$ & 500 & 30\\ \hline
Simulated data for training & PE & 662 & 8,000 & 1 $\times$ 10$^{-8}$ & 500 & 30\\ \hline
Simulated data for testing & PE + CS & 662 & 10,000 & 1 $\times$ 10$^{-8}$ & 500 & 30\\ \hline
Simulated data for testing with 16 energy levels& PE + CS & 1 - 1000 & 10,000*16 & 1 $\times$ 10$^{-8}$ & 500 &30\\ \hline
Simulated data for testing with 6 noise levels & PE + CS &662 & 10,000*6 & 1 $\times$ 10$^{-8}$ & 500 & 5 - 50\\ \hline
Experimental Data & PE + CS & 662 & 50,000 & 1 $\times$ 10$^{-8}$ & 500 & $\approx$ 30\\ \hline

\end{tabular}
\label{tab:all_sim}
\end{adjustbox}
\end{center}
\end{table}

\FloatBarrier

\subsection{Neural network model for Compton and Photonic event detection }

In this subsection, we provide a detailed description of the Multi-Layer Perceptron (MLP) architecture employed for our classification task. The MLP is a feedforward neural network that has shown remarkable effectiveness in modeling complex relationships in various domains. Our choice of an MLP is motivated by its ability to capture intricate patterns and non-linear dependencies in the data.

\begin{figure}[!htb]
  \vspace{-0.5em}
  \begin{center}
        \begin{tikzpicture}[x=2cm,y=1cm]
            \message{^^JNeural network, shifted}
            \readlist\Nnod{7,5,5,5,2} %
            \readlist\Nstr{500,64,64,64,2} %
            \readlist\Cstr{\strut x,a^{(\prev)},a^{(\prev)},a^{(\prev)},y} %
            \def\yshift{0.5} %
            
            \draw[myblue!40,fill=myblue,fill opacity=0.02,rounded corners=2]
              (3.5,-3.5) rectangle++ (1,5.5);

            \message{^^J  Layer}
            \foreachitem \N \in \Nnod{ %
                \def\lay{\Ncnt} %
                \pgfmathsetmacro\prev{int(\Ncnt-1)} %
                \message{\lay,}
                \foreach \i [evaluate={\c=int(\i==\N); \y=\N/2-\i-\c*\yshift;
                             \index=(\i<\N?int(\i):"\Nstr[\lay]");
                             \x=\lay; \n=\nstyle;}] in {1,...,\N}{ %
                    \node[node \n] (N\lay-\i) at (\x,\y) {$\Cstr[\lay]_{\index}$};
            
                    \ifnum\lay>1 %
                        \foreach \j in {1,...,\Nnod[\prev]}{ %
                            \draw[connect,white,line width=1.2] (N\prev-\j) -- (N\lay-\i);
                            \draw[connect] (N\prev-\j) -- (N\lay-\i);
                        }
                    \fi %
                  
                }
                \ifnum\lay<5 %
                    \path (N\lay-\N) --++ (0,1+\yshift) node[midway,scale=1.5] {$\vdots$};
                \fi %
          }
          
          \node[above=5,align=center,mygreen!60!black] at (N1-1.90) {Input\\[-0.2em]layer};
          \node[above=5,align=center,myblue!60!black] at (N3-1.90) {Hidden layers};
          \node[below=160,align=center,myblue!60!black] at (N4-1.90) {t-SNE feature extraction};
          \node[above=10,align=center,myred!60!black] at (N\Nnodlen-1.90) {Output\\[-0.2em]layer};
          
        \end{tikzpicture}
  \end{center}

  \vspace{1em}
  \caption{Neural network architecture of our CoPhNet, The input dimension marked in green have 500 nodes corresponding to the input pulse dimension. We have 3 hidden layers marked in purple each with 64 nodes. For the output layer marked in red, we have two nodes corresponding to the prediction probability of the PE/CS event. The final prediction label is the class with higher probability.
  In addition, we extract latent features at the last hidden layer for t-SNE visualization.
  }
  \label{fig:NNstruct}
\end{figure}
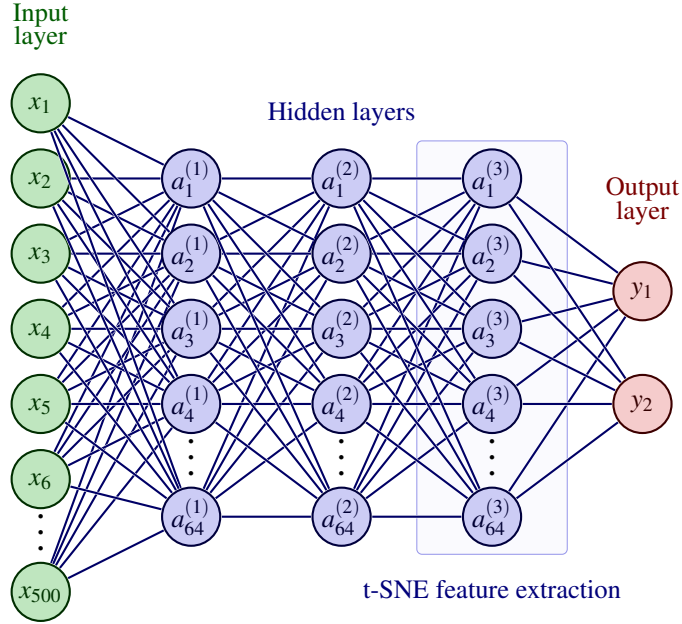

\paragraph{Model Architecture}
The MLP architecture consists of three main components: an input layer, three hidden layers, and an output layer as shown in Fig. \ref{fig:NNstruct}. Each layer is composed of a set of neurons (also known as nodes or units) that perform weighted summations and apply activation functions.

Input Layer: The input layer consists of neurons equal to the dimensionality of our feature vectors (500). Each neuron represents a data point in the input pulse.

Hidden Layers: We employ three hidden layers, each comprising 64 neurons. 
The choice of the number of neurons in these layers was determined through a hyper-parameter tuning process, optimizing model performance.

Output Layer: 
The output layer contains the number of neurons corresponding to the classes in our classification task. We employ a soft-max activation function in the output layer to obtain class probabilities.The final prediction label is the class with higher probability.

\paragraph{Activation Functions}
To introduce non-linearity into the model, we use Rectified Linear Units (ReLU) as the activation function for the neurons in the hidden layers. ReLU has been widely adopted due to its computational efficiency and effectiveness in mitigating the vanishing gradient problem \cite{maier2019gentle}.

\paragraph{Training strategy}
For robust model evaluation, we adopt a 10-fold cross-validation training strategy. The training dataset is divided into ten subsets, and the training process is repeated ten times, each time using a different subset as the validation set while the remaining nine subsets are used for training. This ensures that the model's performance is assessed across a diverse range of data partitions, reducing the risk of over-fitting.

\begin{figure}[!htb] 
  \begin{center}
  \begin{tikzpicture} [node distance=2cm]
        \draw (0, 0) rectangle (12, 1);
        \node at (4.5, 0.5) {Dataset for training(16,000 pulses)};

        \draw (0, 2) rectangle (6, 3);
        \node at (3, 2.5) {Simulated CS
        8,000 pulses};
        
        \draw (6, 2) rectangle (12, 3);
        \node at (9, 2.5) {Simulated PE
        8,000 pulses};

        \draw[->] (6, 1.9) -- (6, 1.1) node[midway, right] {Shuffle};
    \end{tikzpicture}
      
    \begin{tikzpicture}[
        validnode/.style={shape=rectangle, draw=black, line width=0.5,fill=orange!80},
        trainnode/.style={shape=rectangle, draw=black, line width=0.5}
]
        \matrix (M) [matrix of nodes,
        nodes={minimum height = 5mm, minimum width = 8mm, outer sep=0, anchor=center, draw},
        column 1/.style={nodes={draw=none}, minimum width = 4cm},
        row sep=1mm, column sep=-\pgflinewidth, nodes in empty cells,
        e/.style={fill=orange!80}
      ]
      {
        Fold 1 & |[e]| & & & & & & & & & \\
        Fold 2 & & |[e]| & & & & & & & & \\
        Fold 3 & & & |[e]| & & & & & & & \\
        Fold 4 & & & & |[e]| & & & & & & \\
        Fold 5 & & & & & |[e]| & & & & & \\
        Fold 6 & & & & & & |[e]| & & & & \\
        Fold 7 & & & & & & & |[e]| & & & \\
        Fold 8 & & & & & & & & |[e]| & & \\
        Fold 9 & & & & & & & & & |[e]| & \\
        Fold 10 & & & & & & & & & & |[e]| \\
      };

    \matrix [draw,below left] at (7,3) {
      \node [validnode,label=right:validation] {}; \\
      \node [trainnode,label=right:training] {}; \\
    };

    \end{tikzpicture}
    \end{center}
    \vspace{-1em}
    \caption{Data preparation and 10-fold validation illustrations.}
    \label{fig:10fold}
\end{figure}
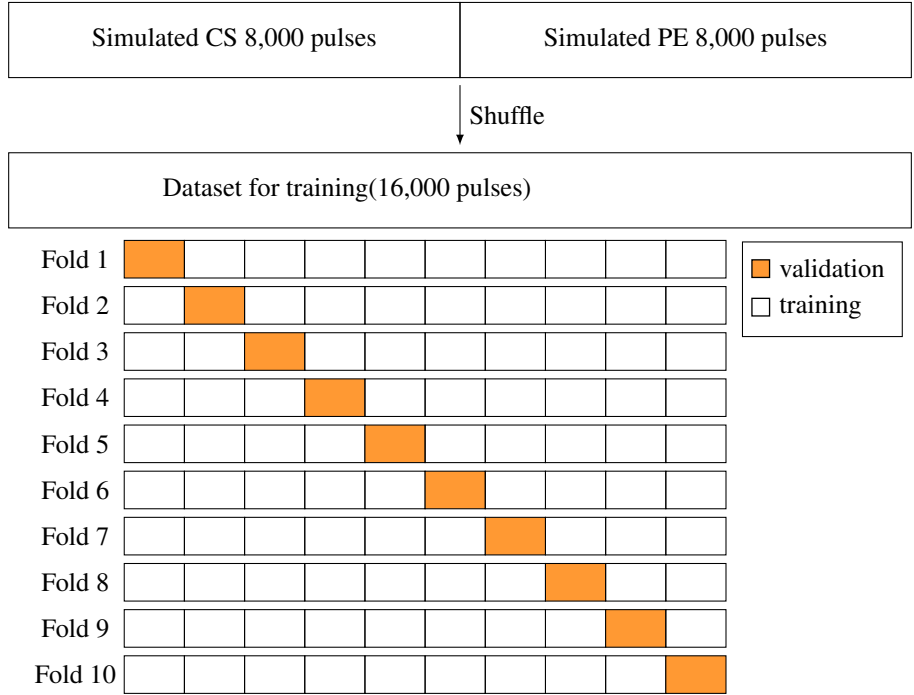

During each fold of cross-validation, we train the MLP network using the back-propagation algorithm and the Adam optimizer while employing the Negative Log-Likelihood Loss ($NLL$) as the loss function Eq. \ref{eq:loss}. 
This loss measures the alignment between the predicted class probabilities and the true class labels, making it suitable for multi-class classification tasks.
\begin{equation}
    NLL(y) = -{\log(p(y))}
\label{eq:loss}
\end{equation}

\paragraph{Network Visualization}
To gain insights into the learned feature representations within our neural network, we employed t-SNE, a powerful dimensionality reduction technique. t-SNE projects high-dimensional feature vectors into a lower-dimensional space while preserving the pairwise similarities between data points. We focused on the feature vectors extracted from the network just before the final classification layer as shown in Fig. \ref{fig:NNstruct}.
 
By doing so, it allows us to visualize and explore the complex relationships and clusters present in the data.

It is crucial to acknowledge that the t-SNE visualization graphs may exhibit variations when regenerated or re-run. 
The primary reason for this variability lies in the stochastic nature of the t-SNE algorithm itself.
t-SNE operates by iteratively optimizing the arrangement of data points in a lower-dimensional space, and during this optimization process, it employs a random initialization of data points. 
This initial randomness, coupled with the inherent non-convex nature of the optimization problem, results in different local minima, making the final projection susceptible to minor perturbations in the input data or the algorithm's parameters. 
Therefore, re-running t-SNE with the same data may yield slightly different layouts or cluster arrangements in the visualization. 
Consequently, it is essential to recognize this variability when interpreting t-SNE visualizations and to consider multiple runs or robustness checks to ensure the reliability of the observed patterns.

We followed the following steps for feature visualization:

\begin{enumerate}
    \item Extracted feature vectors from the neural network for a representative subset of the labeled training set.
    \item Extracted feature vectors for the test set.
    \item Applied t-SNE to reduce the dimensionality of these feature vectors.
    \item Visualize the lower-dimensional representations in two dimensions.
    \item Using the coloring scheme to study the training and testing data distribution.
\end{enumerate}

Concatenating both labeled and unlabeled data provides us with two distinct advantages. 
This way, we can not only leverage the clustering and spatial information to assess the prediction quality for unlabeled data, but also delve into the  distribution shift between the labeled and unlabeled datasets.

\FloatBarrier

\subsection{Evaluation criteria}
The following criteria has been evaluated to validate the efficiency of the CoPhNet model.

\paragraph{Accuracy} 
Accuracy is a fundamental performance metric used to measure the overall correctness of our model's predictions. It is calculated as the ratio of correctly predicted instances to the total number of instances in the dataset. Mathematically, accuracy is defined as:
\begin{equation}
    \text{Accuracy} = \frac{\text{Number of Correct Predictions}}{\text{Total Number of Predictions}} \times 100\%
\end{equation}

Number of correct predictions refers to the count of instances for which the predicted class matches the actual class and the total number of predictions refer to the total number of instances in the dataset subject to prediction.

\paragraph{Confusion matrix} 

The confusion matrix is a detailed assessment tool that breaks down the model's predictions into four categories: true positives (TP), true negatives (TN), false positives (FP), and false negatives (FN). It provides a comprehensive view of the model's performance, especially in scenarios with imbalanced classes. The confusion matrix is structured as shown in Table \ref{tab:confusion_matrix} 

\begin{table}[ht]
\centering
\caption{The layout of confusion matrices.}
\begin{tabular}{>{\centering\arraybackslash}m{3cm}|>{\centering\arraybackslash}m{3cm}|>{\centering\arraybackslash}m{3cm}|}
\cline{2-3}
& \textbf{Predicted Positive} & \textbf{Predicted Negative} \\
\hline
\multicolumn{1}{|>{\centering\arraybackslash}m{3cm}|}{\textbf{Actual Positive}} & True Positive (TP) & False Negative (FN) \\
\hline
\multicolumn{1}{|>{\centering\arraybackslash}m{3cm}|}{\textbf{Actual Negative}} & False Positive (FP) & True Negative (TN) \\
\hline
\end{tabular}
\label{tab:confusion_matrix}
\end{table}

From the confusion matrix, we derive various performance metrics such as precision, recall, F1-score, and specificity, which offer a more nuanced understanding of our model's behavior, particularly in distinguishing between different classes.
These evaluation criteria, accuracy, and the confusion matrix, collectively provide a comprehensive assessment of our model's performance, ensuring that our analysis goes beyond simple accuracy to capture the nuances of its predictions.

\section{Results}
\label{sec:result}

\subsection{Classification performance over simulation dataset }

To evaluate the classification performance of the CoPhNet, we first visualize the training set using a 10-fold cross-validation procedure. We used the combined 16,000 pulse samples and split them according to Fig. \ref{fig:10fold}. For this cross-validation tests, we attained a remarkable 100\% accuracy for all 10 folds, showcasing an unexpected high prediction capability of our CoPhNet. We also achieved 100\% accuracy on the mixed test set, indicating that our model did not overfit on the training set.
To explain this high accuracy, we used the t-SNE to visualize the distribution of all training and testing samples in terms of their latent features extracted from the last hidden layer of our trained CoPhNet. 

\begin{figure}[!htb] 
  \vspace{-0.5em}
  
  \begin{subfigure}[b]{0.475\textwidth}
  \centering
    \includegraphics [width=1\textwidth]{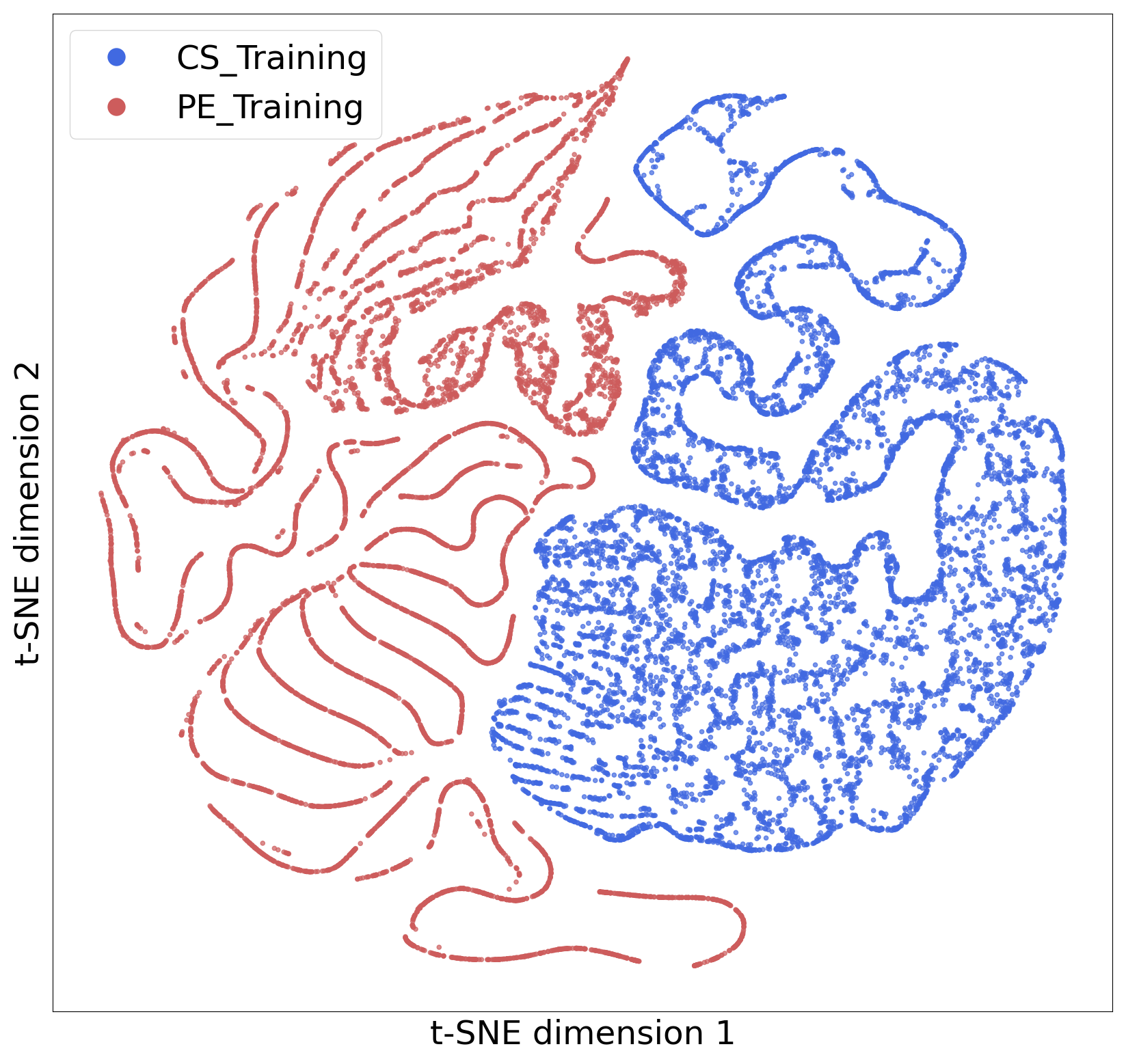}
    \caption{}
  \end{subfigure}
    \vspace{1em}
  \begin{subfigure}[b]{0.475\textwidth}
    \centering
    \includegraphics [width=1.005\textwidth]{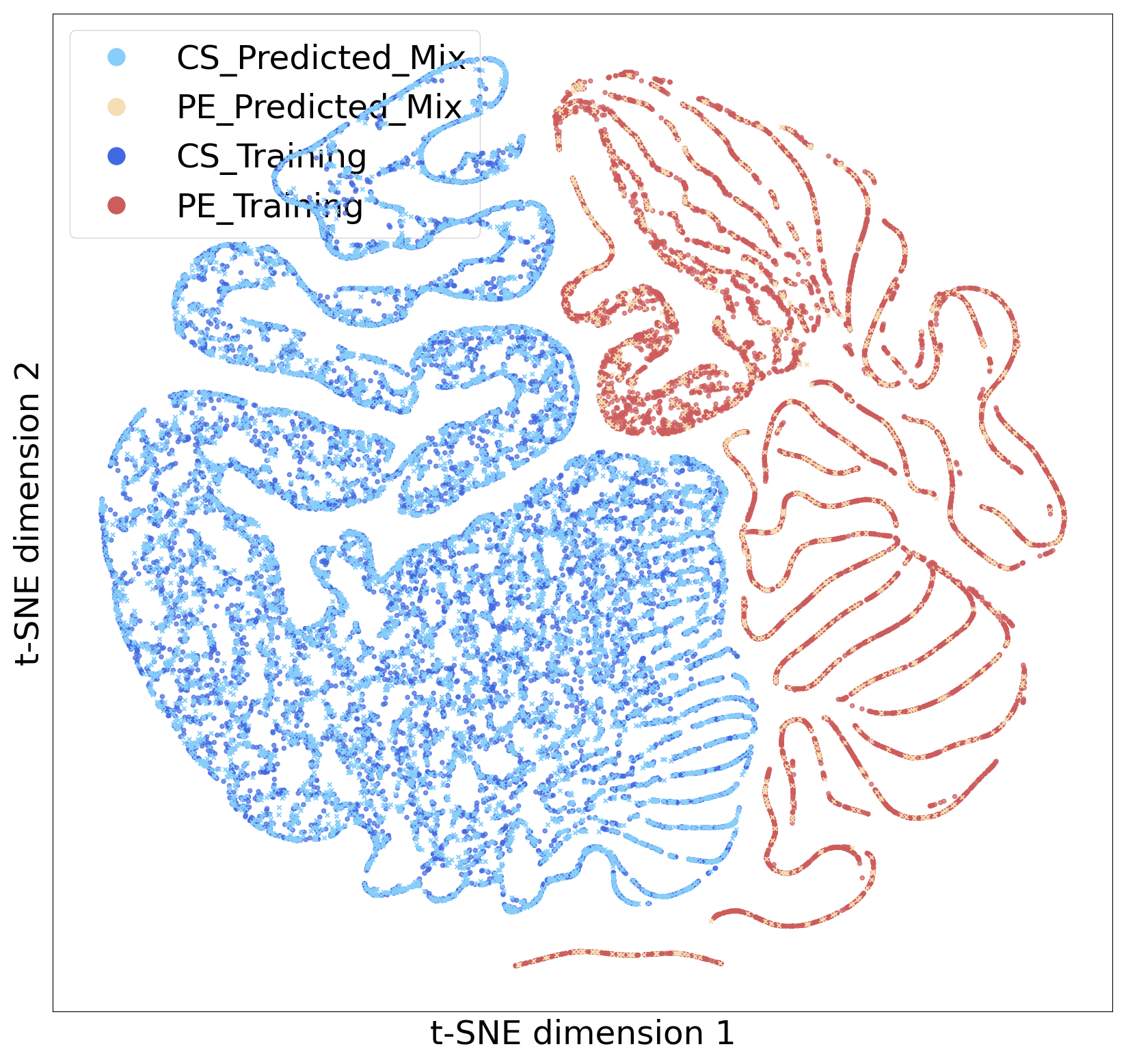}
    \caption{}
  \end{subfigure}

  \vspace{1em}
  \caption{
    t-\uppercase{sne} visualiztion of the training set using latent features extracted from the last FC layer. Different colors indicate the different labels.
  (a) visualization map of the training set using latent features 
  (b) T-sne visualization of the test sets and the training set using latent features.
  }
  \label{fig:trainresult}
\end{figure}
\begin{figure}[!htb] 
  \vspace{-0.5em}   
  
  \begin{subfigure}[!htb]{0.475\textwidth}
  \centering
    \includegraphics [width=1\textwidth]{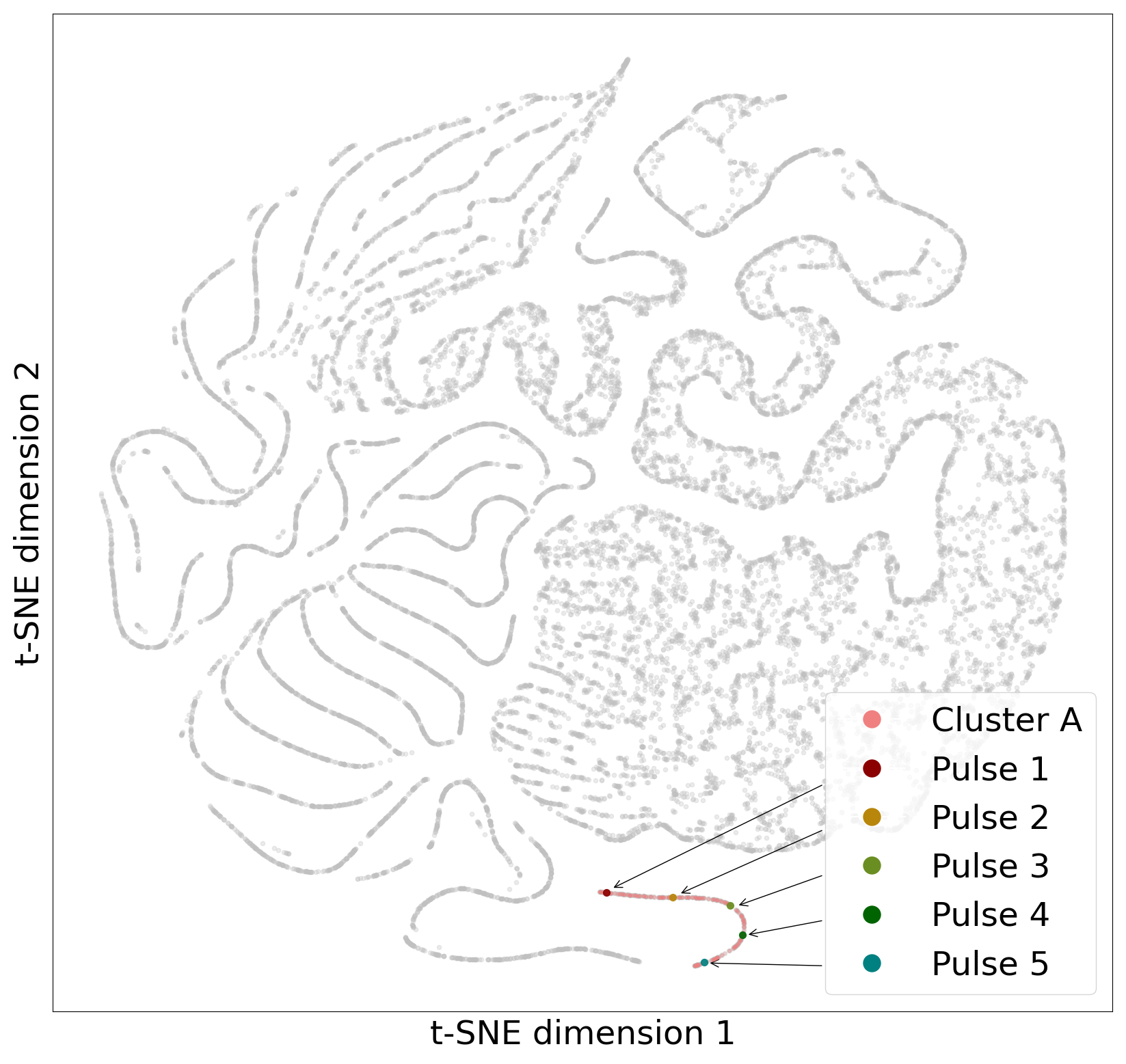}
    \caption{}
  \end{subfigure}
    \vspace{1em}
  \begin{subfigure}[!htb]{0.49\textwidth}
    \centering
    \includegraphics [width=1\textwidth]{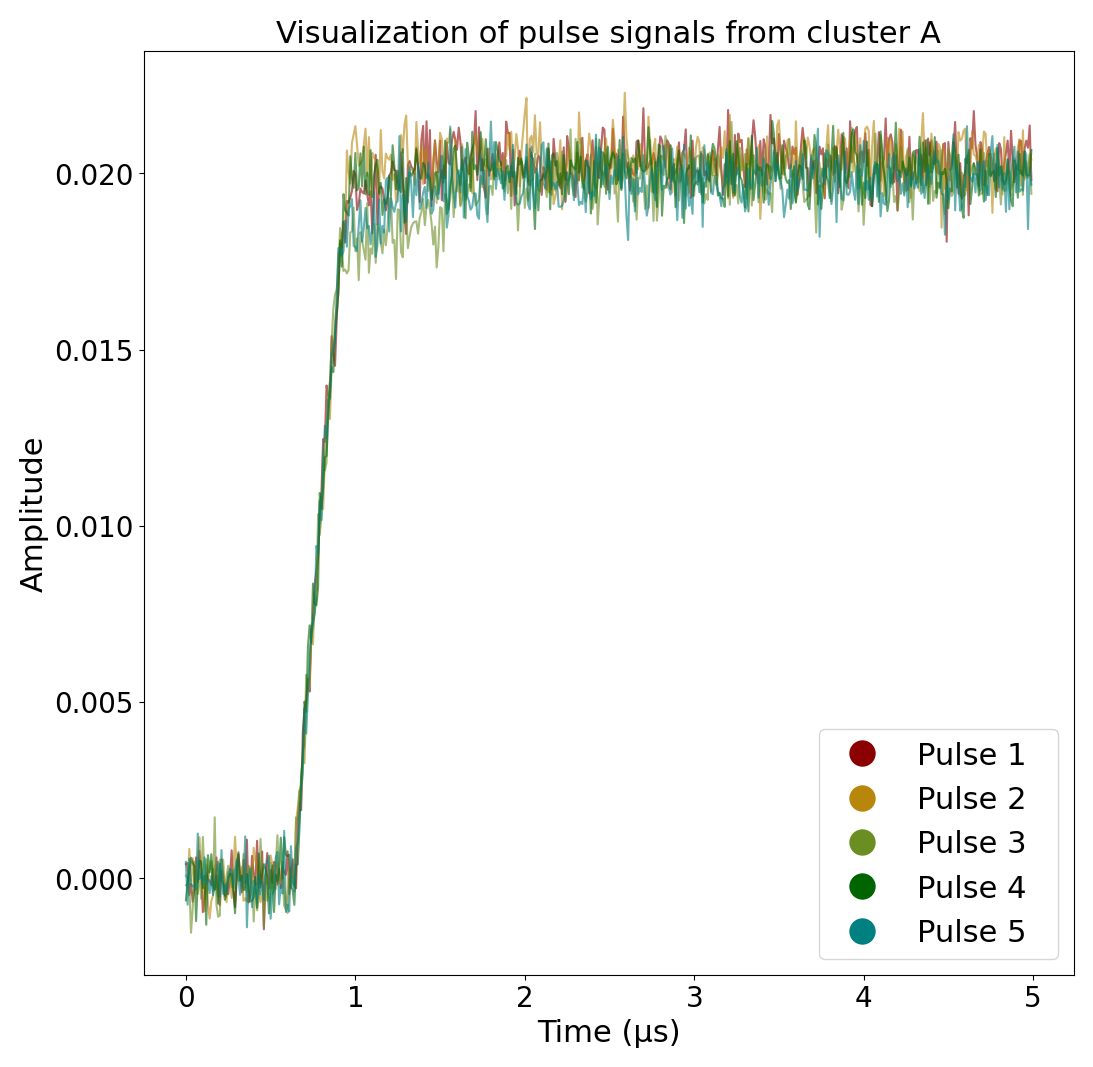}
    \caption{}
  \end{subfigure}

  \vspace{1em}
  \caption{ Commonality of PE pulses within a string-shaped PE cluster. 
  (a) PE cluster and pulse sample selections
  (b) Pulse signals of the PE cluster.}
  \label{fig:traincluster}
\end{figure}

The distribution of the training samples of PE and CS pulse signals are shown in Fig. \ref{fig:trainresult} (a). 
As mentioned early, we visualize the learned feature vectors of all samples as the output of the last hidden layers of the CoPhNet. Within this figure, each data point represents an input pulse, with colors denoting their respective event types—red for PE events and blue for CS events. In t-SNE visualization, the distance between two sample points are corresponding to their similarity. We first observe that the CS samples and PE samples are well separated, indicating our model's effectiveness in classifying both events. Notably, a distinct gap emerges between the clusters associated with PE and CS events. This observation highlights the network's capability in extracting effective features from raw pulse signals and discerning critical differences between these two event categories. Additionally, upon examining the shapes of these clusters, we gain insight into the composition of various pulse shapes within the training set. The red clusters assume string-like island patterns, indicating potential intra-relationships among PE islands. Furthermore, discernible variations are evident among different groups of PE event pulses indicated by different islands, which can also be adeptly distinguished by our classifier. Unlike the PE event signals, the CS event pulses converge into a singular, cohesive cluster. Within this unified structure, numerous smaller string-like clusters are discernible.
Then, we visualize the distribution of the test samples together with the training samples using their latent features in Fig.\ref{fig:trainresult} (b). In this figure, we employ yellow and cyan dots to represent the predicted PE and CS events in addition to the original red and blue dots for training PE and CS signals. A distinct gap emerges between the clusters associated with PE and CS events for both training and testing sets. At the same time, since the training and test sets are generated using the same parameters, their samples shared high similarity within the same event class. This is observable by overlapping of cyan and blue or yellow and red data points.
Since our CoPhNet achieved 100\% accuracy, we used the predicted label for the visualization to simulate the analysis for the unlabeled data.
Proving this method is efficient for visualizing unlabeled data.
Next, to study the commonality of the samples within a cluster, we randomly select five samples from a one-string shaped cluster in the t-SNE map as shown in Fig. \ref{fig:traincluster} (a). The cluster we selected is located at the bottom right of the figure marked in pink. The five samples are marked with 5 different colors corresponding to their pulse visualizations in Fig. \ref{fig:traincluster} (b). The fact that their pulse shapes are highly similar shows the effectiveness of our method.

\FloatBarrier
\subsection{Classification performance over experimental dataset }

\begin{figure}[!htb] 
  \vspace{-0.5em}
  \centering
  \begin{subfigure}[b]{0.8\textwidth}
    \centering
    \includegraphics [width=0.9\textwidth]{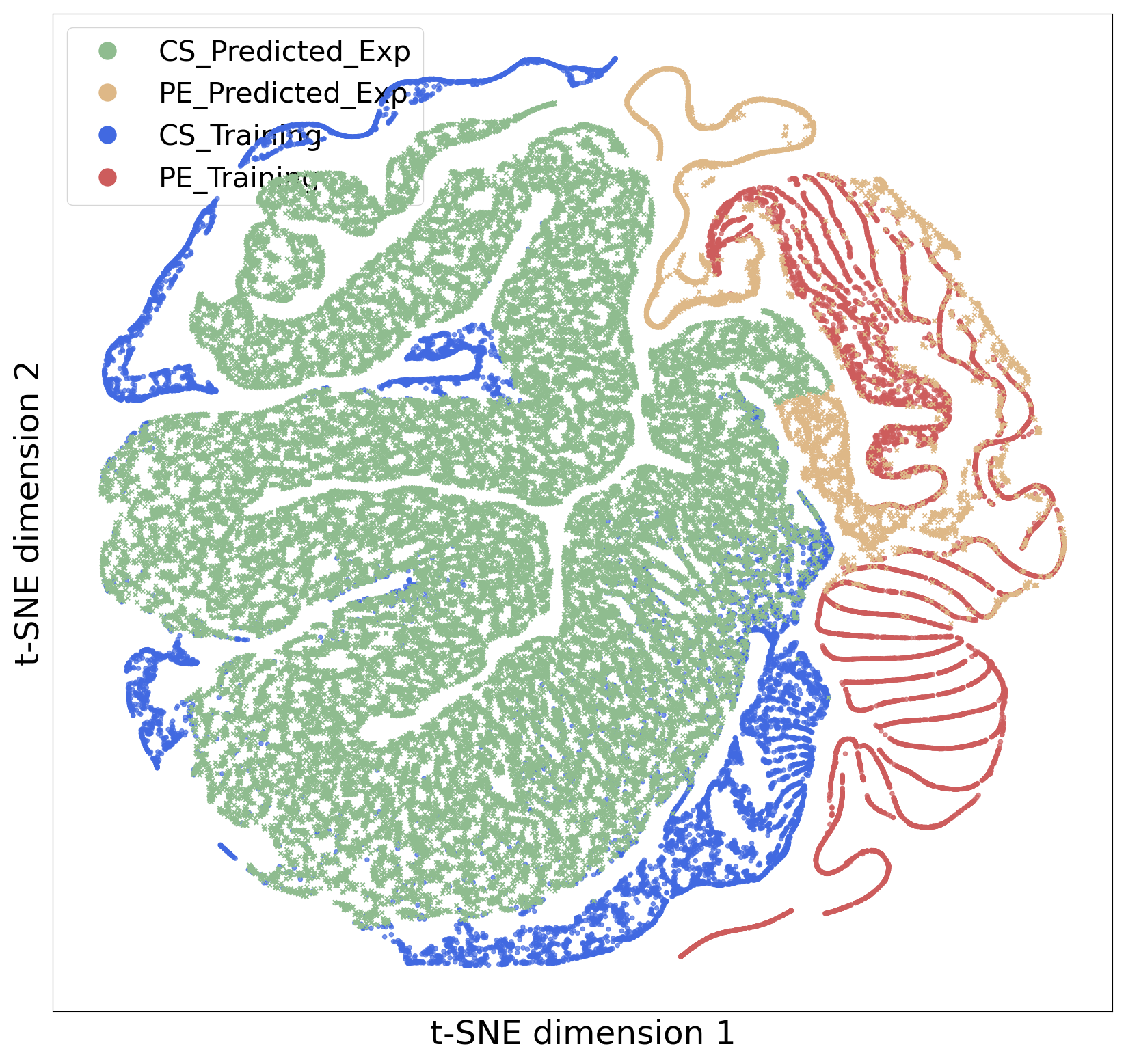}
  \end{subfigure}
  \vspace{1em}
  \caption{Latent feature visualization of the experimental dataset overlaid with the samples of the simulation training set. Different colors indicate sample event types. CS\textunderscore Predicted \textunderscore Exp are the events in the experimental dataset that have been classified as Compton scattered events and PE\textunderscore Predicted\textunderscore Exp are those classified as photoelectric events.}
  \label{fig:exp_tsne}
\end{figure}

One of the challenges we face is the difficulty in evaluating the classification performance of our CoPhNet model over the experimental dataset since we have no access to the true PE/CS event labels of the experimental data. To address this issue,  we present a joint visualization of both the experimental dataset and the training dataset. Fig. \ref{fig:exp_tsne} shows the distribution of the experimental event samples alongside the training samples. Given the substantial size of the experimental dataset (50,000) compared to the training set, a majority of data points in this figure are represented by green and yellow dots, denoting the experimental samples. 
It can be observed that the experimental CS samples (blue dots) and the simulated CS samples (green dots) are all located in the left 3/4 space while these two groups of CS events form their own clusters with several interface junction areas. On the other hand, the experimental (yellow dots) and simulation PE event signals (red dots) are located in the right 1/4 region and they also form different clusters. Together, it shows that the predicted PE/CS events of the experimental datasets are all located near their corresponding PE/CS regions, indicating the high likelihood that their labels are predicted correctly. 
At the same time, unlike Fig.\ref{fig:trainresult} (b), there are little overlapping regions within the same class that show the difference between the simulated and experimental data and our CoPhNet identified those differences. Furthermore, we observe that the boundary between two event classes in experimental samples (green and yellow) exhibits inconsistent alignment with spatial distribution. This is an indicator of performance degradation in these regions.

\FloatBarrier
\subsection{Detection performance on simulated datasets with noise and energy disparity }

As has been discussed previously, the multiple detectors used in an imaging or a detection system may vary in their noise performance because of the intrinsic device and material properties. Also, because of the various radioisotopes used in medical imaging or present in nuclear storage, reactors etc., the energy of the gamma rays encountered by the detector may vary. Because of the above reasons, the prediction accuracy of the CoPhNet trained for a given set of photon energy and noise level, needs to be evaluated for that datasets with parameters that may change unavoidably during the application.
Table \ref{tab:valid_para} introduces these crucial parameters and their variations.

\begin{table}[!htb]
\caption{Generation parameters of simulated datasets. The values in parenthesis indicate the parameters used for generating the training data.}
\setlength{\tabcolsep}{5pt}
\begin{tabular}{p{2.5cm}|lllllllllllllllll}
\hline
SNR (dB)  & 5 & 10 & 15 & 20 & 25 & (30) & 40  & 50  &     &     &     &     &     &                             &     &      \\
\\
Incident gamma energy (keV) & 1 & 10 & 30 & 50 & 60 & 80 & 100 & 150 & 200 & 300 & 400 & 500 & 600 & (662) &700  & 800 & 1000 \\ \hline
\end{tabular}
\label{tab:valid_para}
\vspace{1em}
\end{table}

\begin{figure}[!htb] 
  \vspace{-0.5em}
    \centering
  \begin{subfigure}[!htb]{0.46\textwidth}
    \centering
    \includegraphics [width=1\textwidth]{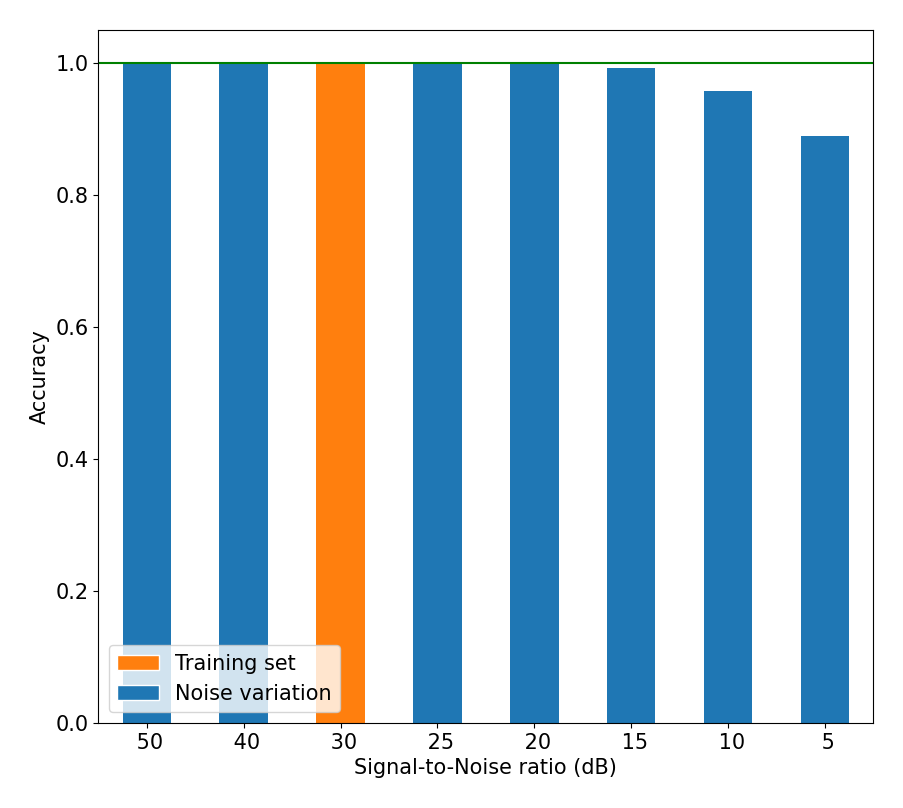}
    \caption{}
  \end{subfigure}
  \hspace{1em}
  \begin{subfigure}[!htb]{0.46\textwidth}
  \centering
    \includegraphics [width=1\textwidth]{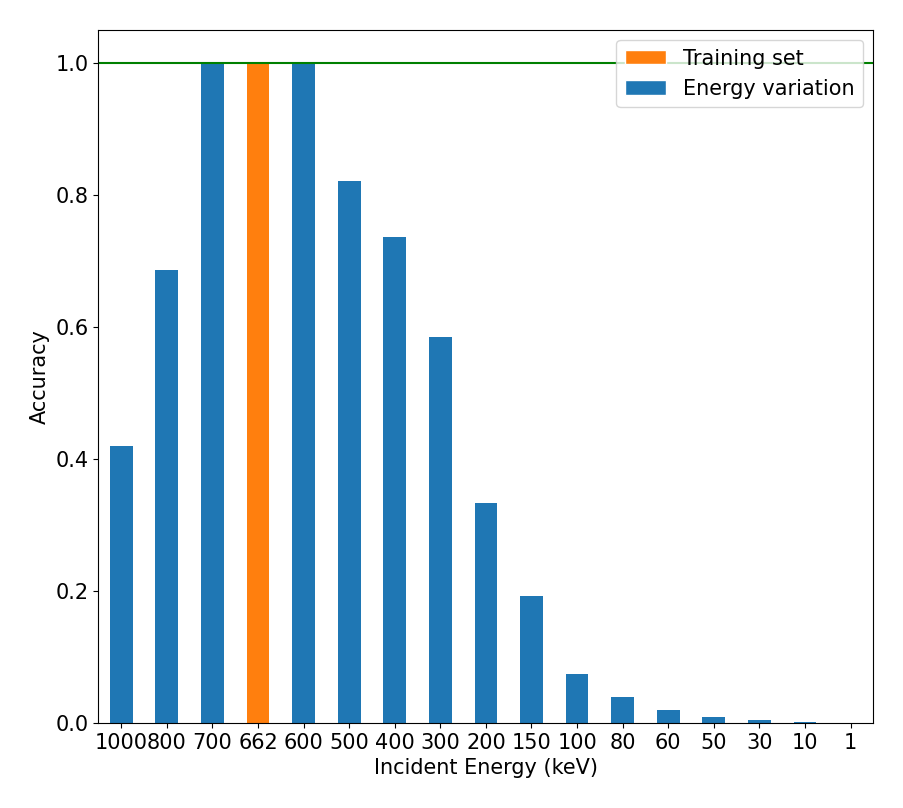}
    \caption{}   
  \end{subfigure}

  \caption{
  (a) Accuracy comparison over different noise levels.
  (b) Accuracy comparison over different incident energies.
  }
  \label{fig:energy_noise_acc}
\end{figure}

\subsubsection{Classification performance over noise level variation}

In order to evaluate the performance of our CoPhNet, trained using dataset comprising pulses with a specific noise level, against the variation in detector noise level, we have validated the CoPhNet for datasets comprising pulses with different signal-to-noise ratios (SNRs). The noise has been added to the simulated signal as additive white Gaussian noise. The SNR was set to 30 dB for the training data to resemble the SNR of the pulses obtained from our experimental setup. Test data with SNR varying within 5 - 50 dB were generated. Figure \ref{fig:energy_noise_acc} (a) shows the variation of prediction accuracy as a function of the noise level. An SNR of 50 dB implies very clean pulses (low noise) whereas a SNR of 5 dB implies a noisy signal. A noisy signal results in uncertainty in the calculated pulse-heights or rise-times. It can be observed from Fig. \ref{fig:energy_noise_acc} (a) that the accuracy is mostly 1 for SNR down to 15 dB and deteriorates to $\approx$0.85 for the dataset with SNR $=$ 5 dB. It can be inferred from the results shown in the figure that the CoPhNet is quite resilient against variation in the detector noise levels and did not show any dependence on the SNR when it is higher than that used for the training dataset. A nominal dependency on the noise level for input datasets with SNRs much lower than that of the training dataset has been observed.

\begin{figure}[!ht] 
  \vspace{-1em}
  \begin{subfigure}[!htb]{0.45\textwidth}
    \centering
    \includegraphics [width=1\textwidth]{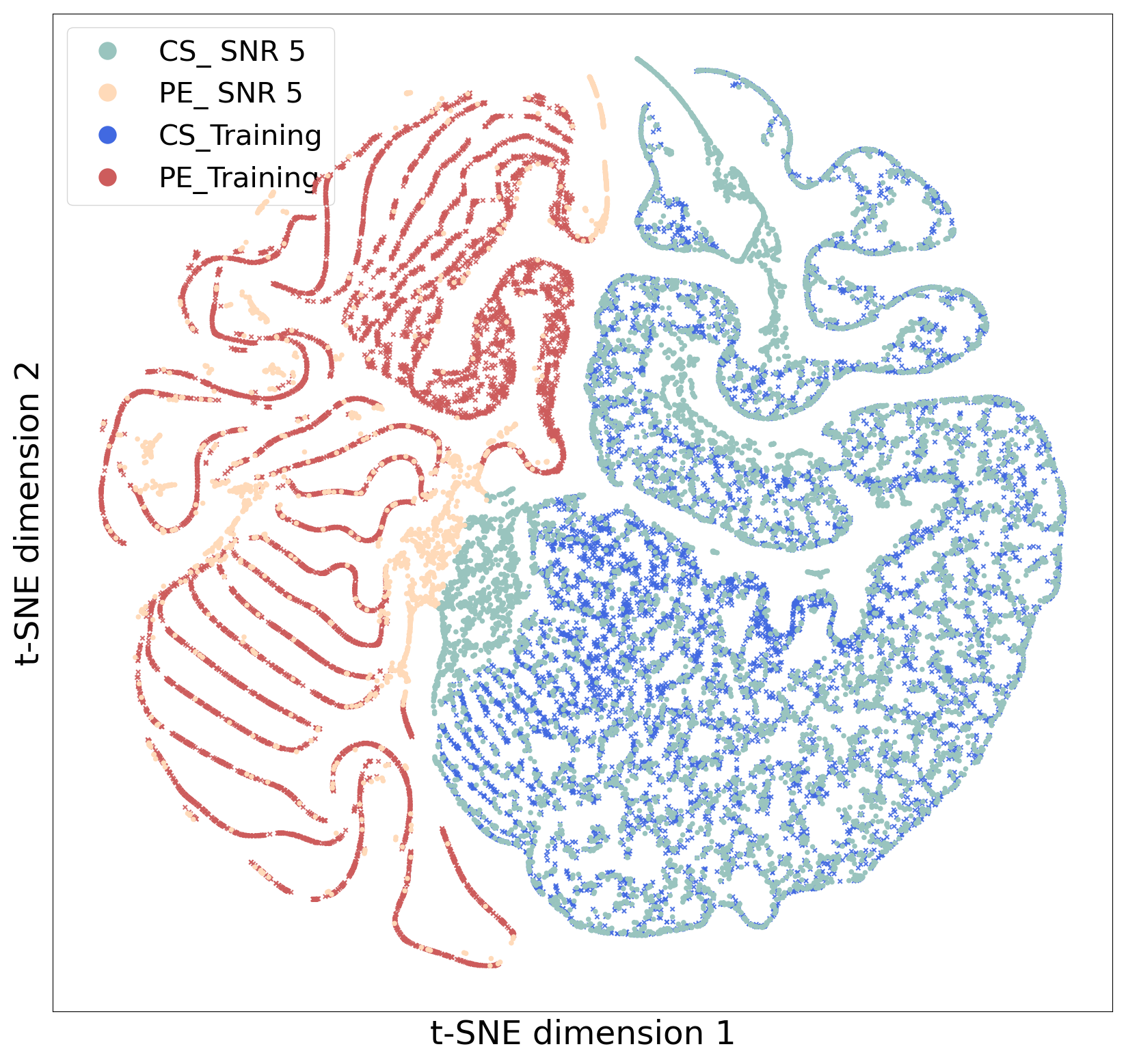}
    \caption{}
  \end{subfigure}
    \hspace{1.8em}
  \begin{subfigure}[!htb]{0.45\textwidth}
  \centering
    \includegraphics [width=1\textwidth]{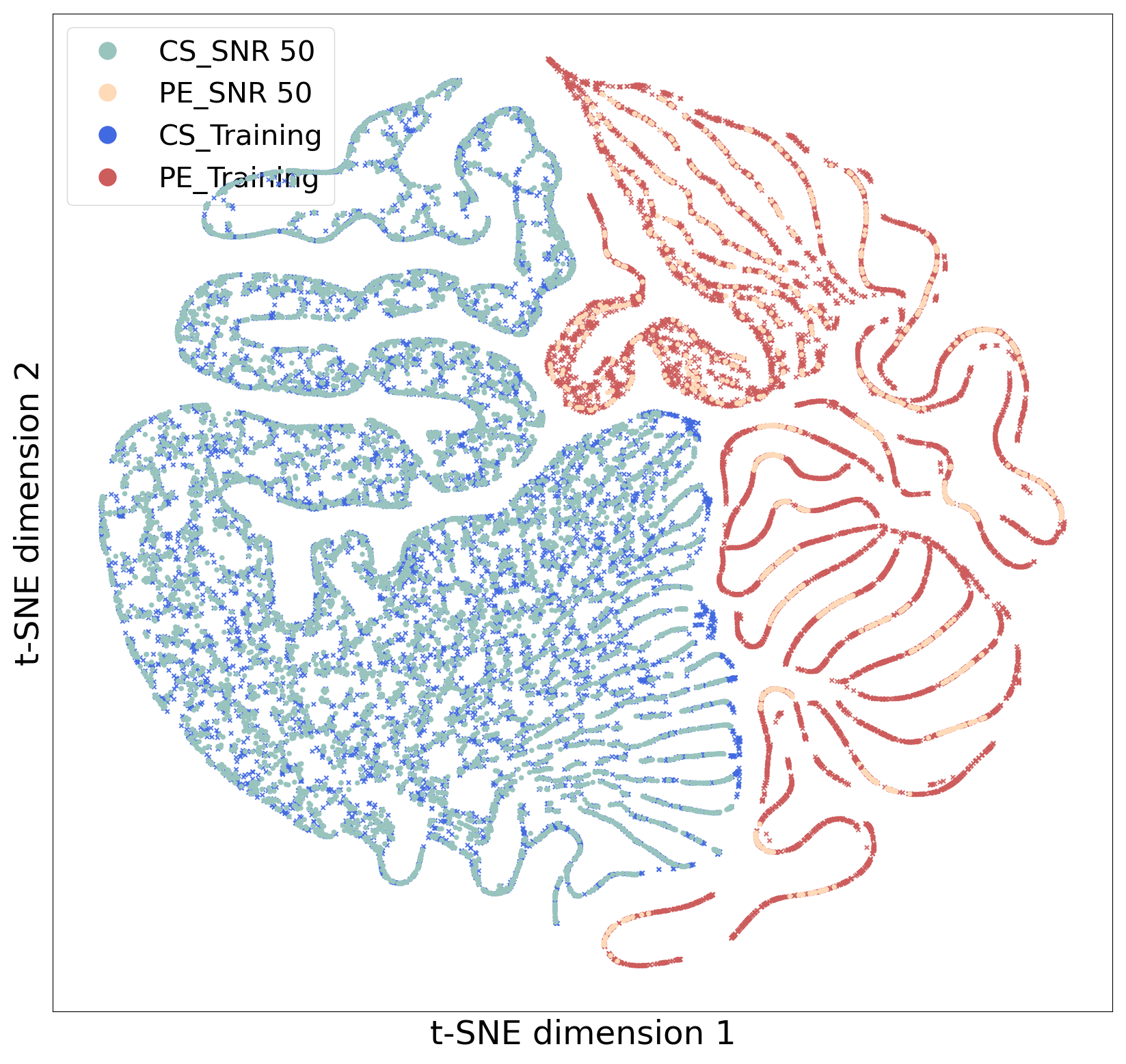}
    \caption{}   
  \end{subfigure}

  \vspace{.5em}
    \begin{subfigure}[!htb]{0.45\textwidth}
    \centering
    \includegraphics [width=1\textwidth]{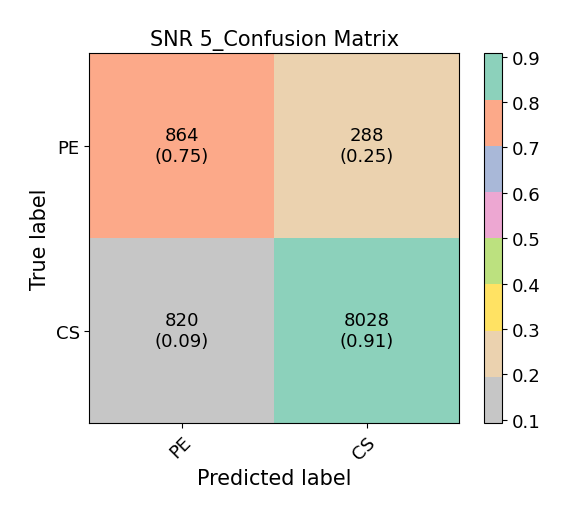}
    \caption{}
  \end{subfigure}
      \hspace{1.8em}
\begin{subfigure}[!htb]{0.45\textwidth}
    \centering
    \includegraphics [width=1\textwidth]{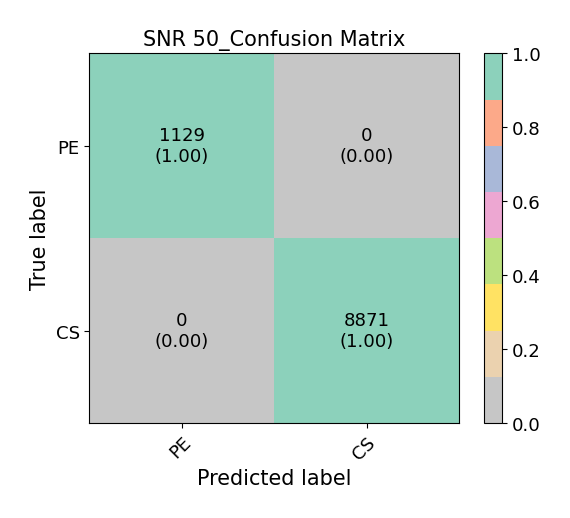}
    \caption{}
  \end{subfigure}
  
  \vspace{-.5em} 
  \caption{
    (a)\&(b): Latent features visualization for datasets with different noise levels overlaid with the samples of the training set.
    Different colors indicate different labels.
    (c)\&(d): Confusion matrix for datasets with different noise levels. Different colors indicate the number of samples.
  (a) SNR 5 latent features.
  (b) SNR 50 latent features.   
  (c) SNR 5 Confusion matrix.
  (d) SNR 50 Confusion matrix.
  }
  \label{fig:noise_tsne}
\end{figure}

As mentioned earlier, the inconsistent color and shape boundary suggests performance drop in the region. We observed this phenomena in Fig. \ref{fig:exp_tsne}. Similarly, the boundary between CS/PE event feature visualization (green \& yellow) shown in Fig.\ref{fig:noise_tsne} (a) indicates the mis-classification. This is verified by the confusion matrix shown in Fig. \ref{fig:noise_tsne} (c). At the same time, similarity of simulated samples are recognized by our CoPhNet thus the overlapping of the same class samples on visualization are as same as the earlier mentioned mixed dataset shown in both Fig.\ref{fig:noise_tsne} (a) \& (b). Our CoPhNet demonstrated 100 \% accuracy for the SNR 50 (dB) dataset as is confirmed by the confusion plot in Fig.\ref{fig:noise_tsne} (d). Comparing Fig.\ref{fig:noise_tsne} (b) \& Fig.\ref{fig:trainresult} (b), we noticed that with less noise in the signal, our CoPhNet performs better in identifying the difference between the CS and PE events. 
Unlike Fig.\ref{fig:trainresult} (b) where same class samples are evenly distributed, the samples that are located at the event boundaries in Fig.\ref{fig:noise_tsne} (b) are mostly blue and red which are training samples with a noisier signal pulses(30 dB). The green and yellow samples (50 dB) are located away from the class boundary.

\subsubsection{Classification performance over incident energy variation}

\begin{figure}[!ht] 
 \centering
  \vspace{-0.5em}
  \begin{subfigure}[c]{0.48\textwidth}
    \includegraphics [width=1\textwidth]{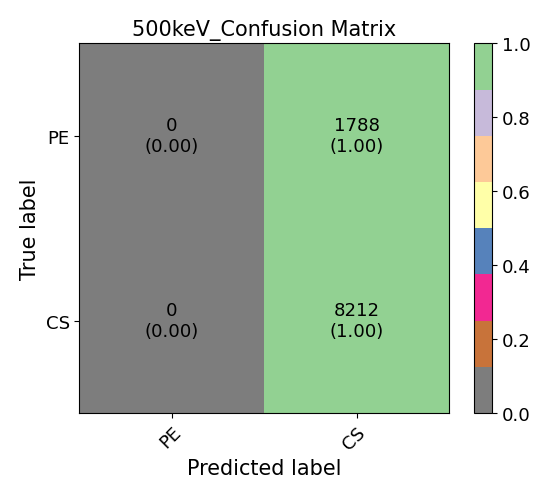}
    \caption{}
  \end{subfigure}
  \begin{subfigure}[c]{0.48\textwidth}
    \includegraphics [width=1\textwidth]{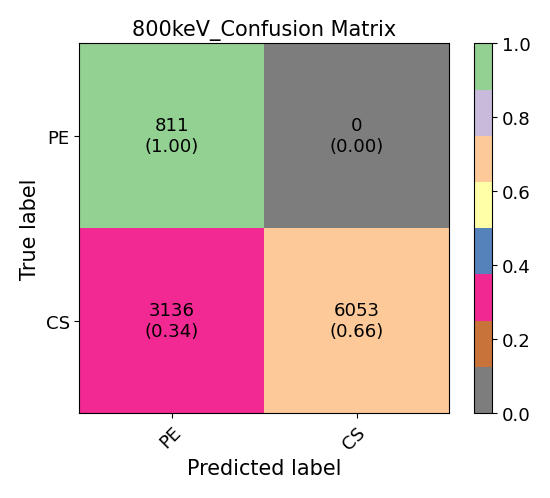}
    \caption{}   
  \end{subfigure}
    \vspace{1em}
  \caption{
  Confusion matrices for datasets with different incident gamma energies.
  In each block, the top number indicates the number of samples in this class.
  The values in parenthesis indicate the percentage of the sample in this block over true label samples.
  (a) 500 keV
  (b) 800 keV.
  }
  \label{fig:ene_CM_500_800}
\end{figure}

To assess the impact of changing incident photon energies on the CoPhNet model trained with dataset comprising pulses generated in a CZTS detector facing 662 keV gamma photons, the model was evaluated for its performance over signals generated for interactions with photon energies varied between 1 keV to 1 MeV. 
We conducted accuracy calculations across all the energy levels and present the results in Fig. \ref{fig:energy_noise_acc} (b). Unlike in the case of varying noise-level dataset, variations in incident energy may significantly impact the distribution of the two types of events. Compton scattering occurs primarily in the medium (above 10 keV) to high energy range (several MeVs) of gamma photons while photoelectric events are more probable for lower energy. This distribution shift introduces considerable instability in our model's performance as is evident from the strong dependency of the prediction accuracy on the incident gamma energy shown in Fig.\ref{fig:energy_noise_acc}  (b). 
Impressively, our model achieved a remarkable accuracy of 99.97\% at 600 keV and 99.94\% at 700 keV. Those observations demonstrate that although the prediction accuracy diminishes for large energy variation, the CoPhNet is robust against minor energy shifts occurring due to other external factors such as scattering from tissues or shielding materials. 
For two extreme incident energies (1 keV and 1 MeV),we examined their confusion matrix (available at supplementary Fig. S1 (a) \&(b)). Every sample from the 1 keV incident energy dataset has a true label as a PE event, yet our model classifies all of them as CS events. This intriguing outcome yields an accuracy of 0\%. This brings us the question that at which incident energy our CoPhNet starts miss-classifying the entire class. We also draw conclusions that the accuracy score cannot accurately capture the performance of our CoPhNet. In the Fig. \ref{fig:energy_noise_acc}  (b), it seems our CoPhNet still holds decent performance at 800 and 500 keV. But after we examined the confusion matrices, we noticed that the single class accuracy is as low as 0\% as shown in Fig. \ref{fig:ene_CM_500_800} (a) top left block.

Due to the distribution shift of the proportion between two event classes caused by the difference in incident gamma energy, the PE events only takes 17.88\% of the 500 keV dataset and all of PE events are classified as CS pulses, which yields 0\% class accuracy in PE class prediction as shown in the Fig. \ref{fig:ene_CM_500_800} (a) top row. However, the overall accuracy is still 82.12\% (the sum of diagonal blocks divided by the sum of all blocks). At the same time, for the 800 keV dataset (Fig. \ref{fig:ene_CM_500_800} (b)), the accuracy over the CS class is only 66\% as shown in the bottom right block. Yet due to the smaller proportion of the PE samples (8.11\%), the overall accuracy is 68.64\%.

An intriguing pattern emerged when we compared all the confusion matrices in Fig. \ref{fig:ene_CM_500_800} and supplementary Fig. S1. Specifically, for incident energies lower than that of our training set, PE events tend to be misclassified as CS events, and conversely, for incident energies higher than the training set, CS events tend to be misclassified as PE events. Notably, this pattern persists consistently across all energy levels. 

A plausible explanation of the above-mentioned observation is as follows. In an ideal situation, where there is no charge-trapping in the detector, the CS events always register lower energies compared to the PE events for an incident gamma photon energy. In fact, the energy deposited by the 662 keV gamma photons in the CS events ranges from 0 to $\approx$479 keV, assuming a photopeak (full energy peak in the pulse height spectrum) at 662 keV. The observation that the PE events are misclassified as CS events for incident energies lower than that of the training dataset and CS events are misclassified as PE events for incident energies lower than that of the training dataset, implies that CoPhNet model gives a strong weightage to features of the sample pulses that are relevant to incident energy.

\begin{figure}[!ht] 
  \begin{subfigure}[!htb]{0.49\textwidth}
    \includegraphics [width=1\textwidth]{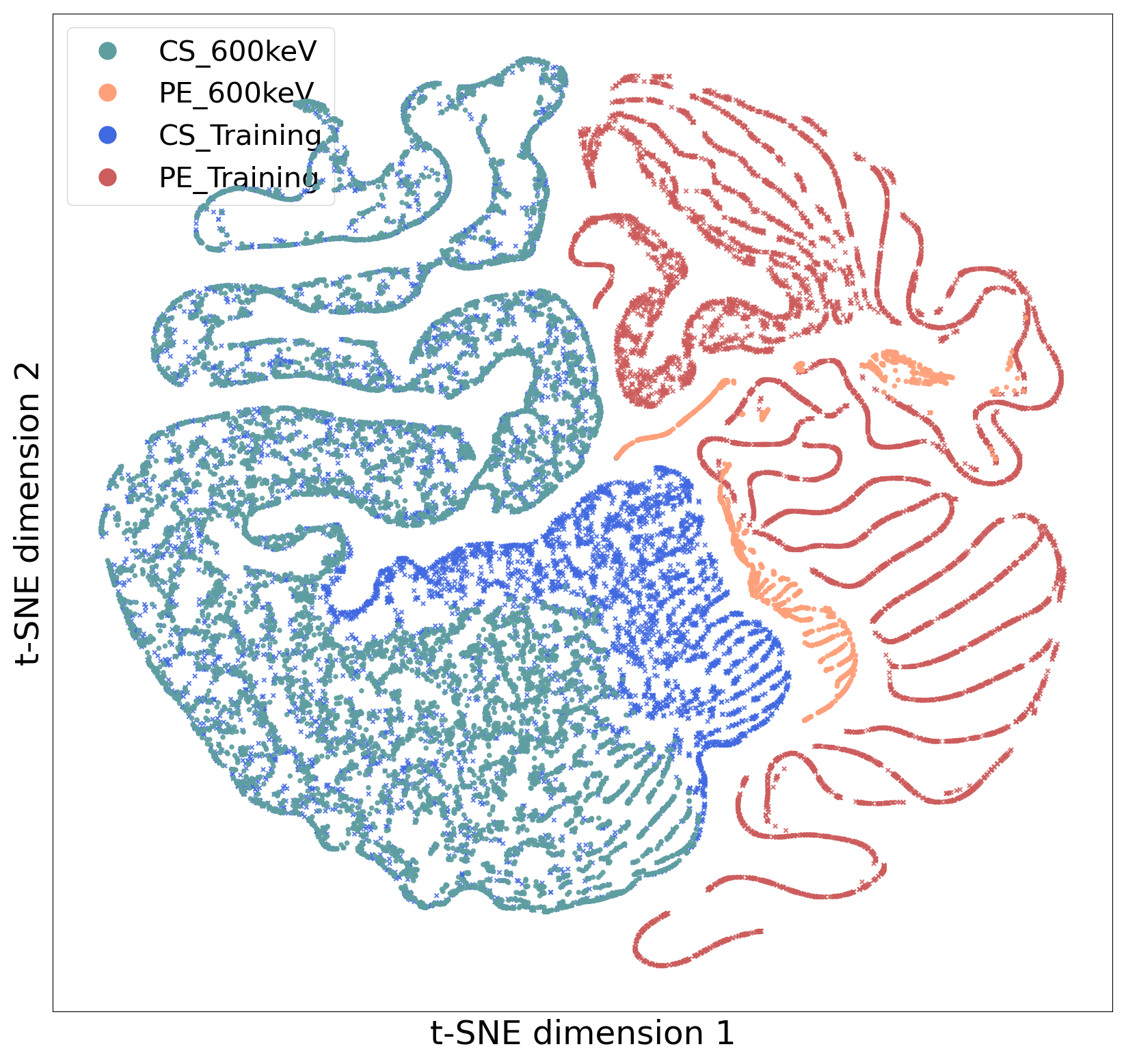}
    \caption{}
  \end{subfigure}
  \hspace{1.1em}
  \begin{subfigure}[!htb]{0.49\textwidth}
  
    \includegraphics [width=1\textwidth]{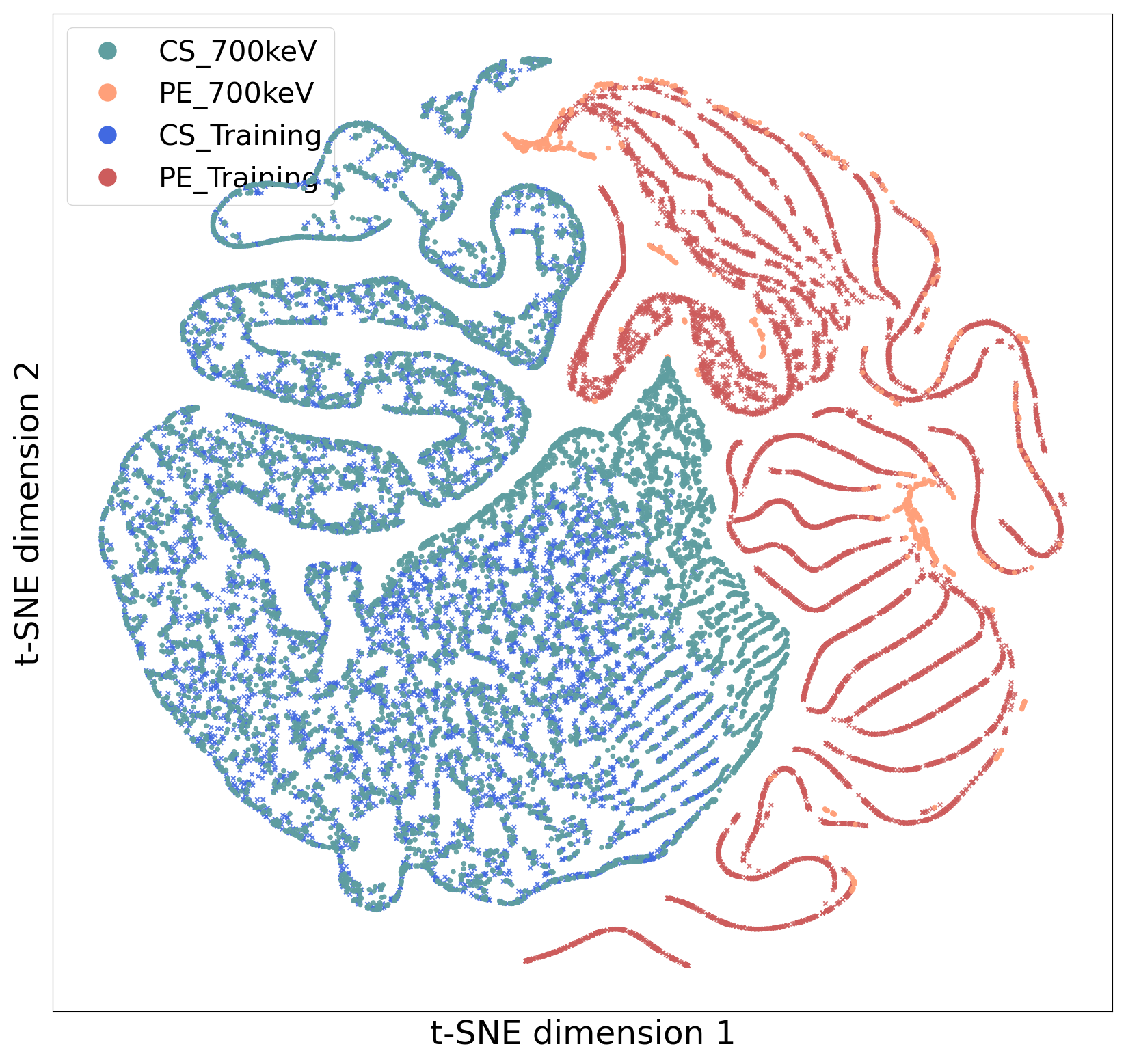}
    \caption{}   
  \end{subfigure}
  
    \centering
  \vspace{1em}
  \caption{
  Latent features visualization for datasets with different incident energies overlaid with the samples of the training set.
  Different colors indicate different labels.
  (a) 600 keV
  (b) 700 keV
  }
  \label{fig:ene_600_700tsne}
\end{figure}

The robustness of our CoPhNet under small incident energy shifts (around 50 keV) allows us to analyze the latent feature visualization of the 600 keV \& 700keV in Fig. \ref{fig:ene_600_700tsne}. In both figures, the fact that different event samples are well separated with distinguished gaps between them is indication of the CoPhNet performance. At the same time, unlike the noise ratio, our model successfully recognizes the difference between the different incident energies. Similar to the Fig.\ref{fig:exp_tsne}, the target incident energy samples have little overlapping regions within the same class of the training samples. In Fig.\ref{fig:ene_600_700tsne} (a), we noticed that the 600 keV PE samples (blue) are very close to the training 662 keV CS samples(yellow). This suggests the likeness of their sample pulse according to the properties of the t-SNE visualization. It also applies to the training 662 keV PE samples (red) and 700 keV CS samples (green) in Fig.\ref{fig:ene_600_700tsne} (b).
Further supporting the pattern we mentioned early that low energy PE can be mis-classified as CS and vice versa for high incident energy.
\FloatBarrier

\section{Conclusion}%
The primary objective of the study is to develop a deep learning enabled gamma photon detection system that can distinguish the output signals produced by photoelectric interactions from those generated by the Compton absorption. Such capabilities is highly sought after for our semiconductor detection system with various applications including medical imaging and radiation monitoring in nuclear reactor sites. 

In this article we developed and trained a deep learning model and explored its performance and robustness in classification of pulse signals generated by gamma rays in CdZnTeSe semiconductor detectors. The systematic investigation took various aspects, such as the effect of change in detector noise levels and photon energies, in consideration while investigating the prediction accuracy of the model. Our model was trained using simulated datasets comprising detector pulses generated when a CdZnTeSe detector interacts with 662 keV gamma photons through photoelectric absorption or Compton scattering. The essential detection features such as material/device properties, readout electronics specifications, and effects of charge trapping have been matched to a real CdZnTeSe detector developed in our laboratory to obtain realistic pulses. Separate datasets comprising either pure photoelectric events or Compton scattered events were used so that the model can learn the pulse features strictly specific to the type of interaction. The model was validated with a mixed dataset comprising randomly distributed detector pulses generated from both the types of interaction. The model remarkably exhibited a 100\% accuracy in predicting the type of interaction. The accuracy was maintained even when the model was validated with datasets generated by altering the detector noise level. The training data simulated with a signal-to-noise ratio of 30 dB did not show any change in accuracy for the datasets generated with higher and immediate lower SNRs. For datasets with very noisy signals (SNR = 5 dB) the prediction accuracy lowered to $\approx$90\%.

An accuracy close to 100\% was achieved when validated with datasets comprising detector signals generated due to the interactions with gamma photons of 600 and 700 keV (a few tens of keVs above and below the training energy of 662 keV). We also found that our model's performance is robust to a reasonable degree of pulse energy variation. 
Beyond these energies the accuracy of the DL model however demonstrated strong dependency on the energy of the gamma photons used to generate the training dataset. Notably, for gamma photon energies of 1 keV, all samples from the simulated dataset were PE events, yet our model classified them uniformly as CS events, resulting in a 0\% accuracy. The accuracy went down to 40\% when validated for the 1 MeV dataset. The observations indicate that the DL model learns the pulse-height related features to distinguish between CS and PE events which mistakes any substantially low energy events as CS events. However, it has also been inferred that the model learns additional features possibly, pulse-height - risetime correlations, leading to the 100\% prediction accuracy when validated for datasets comprising pulses with similar energies to the training dataset. 

Our CoPhNet has also been validated using an experimental dataset from our CZTS detector illuminated to 662 keV gamma rays. Since evaluating the classification performance of the neural network model over the experimental dataset is challenging as there is no access to the true PE/CS event labels in the experimental dataset, we have resorted to the latent feature visualization, which showed that the test samples with PE/CS labels are located near those training samples with same labels, indicating the high likelihood that their labels were predicted correctly. 

In summary, our study demonstrates the robustness of our CoPhNet in handling noise levels and its adaptability to variations in incident energy, upholding its resiliency against inevitable variations in parameters during real applications. While challenges exist, our findings provide valuable insights for signal processing applications and underline the need for further investigation into the model's behavior under diverse conditions.
Our research opens avenues for future exploration, including refining our model to address energy level discrepancies and exploring strategies to enhance its adaptability to a wider range of signal variations.

\section{Data and Code Availability}

The datasets generated in this work and code are available from the corresponding author upon reasonable request.

\section{Contribution}
Conceptualization, J.H. and K.M.; methodology, Q. L, S.C., J.H., K.M.; software, J.H., Q.L., S.C.; resources, J.H. and K.M.; writing--original draft preparation, S.C., Q.L., J.H.; writing--review and editing, J.H.; visualization, Q.L., S.C.; supervision, J.H. and K.M.;  funding acquisition, J.H. and K.M.

\section*{Acknowledgement}
The research reported in this work was supported in part by South Carolina EPSCOR GEAR CRP grant 23-GC02 supported by the National Science Foundation under the grant 1655740. The views, perspectives, and content do not necessarily represent the official views of the NSF. Krishna C. Mandal wants to acknowledge the support from the Health Sciences Center (HSC) Seed Grant, Prisma Health, under Grant 10011863; and the Transformative Research Seed Grant Initiative Award under Grant 80004827.

\bibliographystyle{unsrt}  
\bibliography{references}

\end{document}